\documentclass[11pt,paper]{article}
\pdfoutput=1

\usepackage{jheppub}
\usepackage[usenames,dvipsnames,table]{xcolor}
\usepackage{graphicx,amsmath,amssymb,multirow,array,bm,mathrsfs}
\usepackage{epsf,amsfonts,slashed}
\usepackage[numbers,sort&compress]{natbib}

\usepackage{hyperref}

\def\d{{\rm d}}
\def\sech{\operatorname{sech}}
\def\erfc{\operatorname{erfc}}

\preprint{YITP-SB-16-15}
\title{
The holographic dual of a Riemann problem in a large number of dimensions
}
\author{Christopher P. Herzog$^*$, Michael Spillane$^*$, and Amos Yarom$^\dagger$}

\affiliation{
$*$ C.~N.~Yang Institute for Theoretical Physics, Department of Physics and Astronomy \\
Stony Brook University, Stony Brook, NY  11794, USA
\vskip 0.05in
$\dagger$ Department of Physics, Technion, Haifa 32000, Israel
}

\abstract{
We study properties of a non equilibrium steady state generated when two heat baths are initially in contact with one another. 
The dynamics of the system we study are governed by holographic duality in a large number of dimensions.  
We discuss the ``phase diagram'' associated with the steady state, the dual, dynamical, black hole description of this problem, and its relation to the fluid/gravity correspondence.
}
\begin{document}
\maketitle

%********************************************************
\section{Introduction}

The Riemann problem may provide a relatively simple setting in which to study the non-equilibrium physics of quantum field theory.
The problem asks for the time evolution of piece wise constant initial conditions with a single discontinuity in the presence of some number of conservation laws, for example of energy, momentum, mass, or charge.  In our case, we consider a fluid phase of a conformal field theory (CFT) with an initial planar interface, where the energy density jumps from $e_L$ on the left of the interface to $e_R$ on its right. We also allow for  a discontinuity in the center of mass velocity of the  fluid across the interface.

For simplicity, we will make a number of further restrictions.  We assume a conformal field theory that has a dual gravity description via the AdS/CFT correspondence. A priori, this will allow us to study the system beyond the hydrodynamic limit. We also take the limit that the number of spatial dimensions $d$ is very large.   In this limit, we find that the system is described by two conservation equations
\begin{equation}
\label{E:Tmn}
	\partial_t e - \partial_\zeta^2 e = - \partial_\zeta j \ ,
	\qquad
	\partial_t j - \partial_\zeta^2 j = -\partial_\zeta \left(\frac{j^2}{e}+e\right) \,.
\end{equation}
where $e$ is, up to gradient corrections, the energy density and $j$ the energy current.  These equations are a special case of equations derived in ref.\ \cite{Emparan:2015gva}.
In these variables the Riemann problem amounts to a determination of $e$ and $j$ given an initial configuration of the form
\begin{equation}
\label{E:ini}
	(e,j) = \begin{cases} (e_L,j_L) & z<0 \\ (e_R,j_R) & z>0 \end{cases}\,.
\end{equation}
By choosing an appropriate reference frame, we may set $j_L=0$ without loss of generality.

\begin{figure}
\centering
\includegraphics[width=3in]{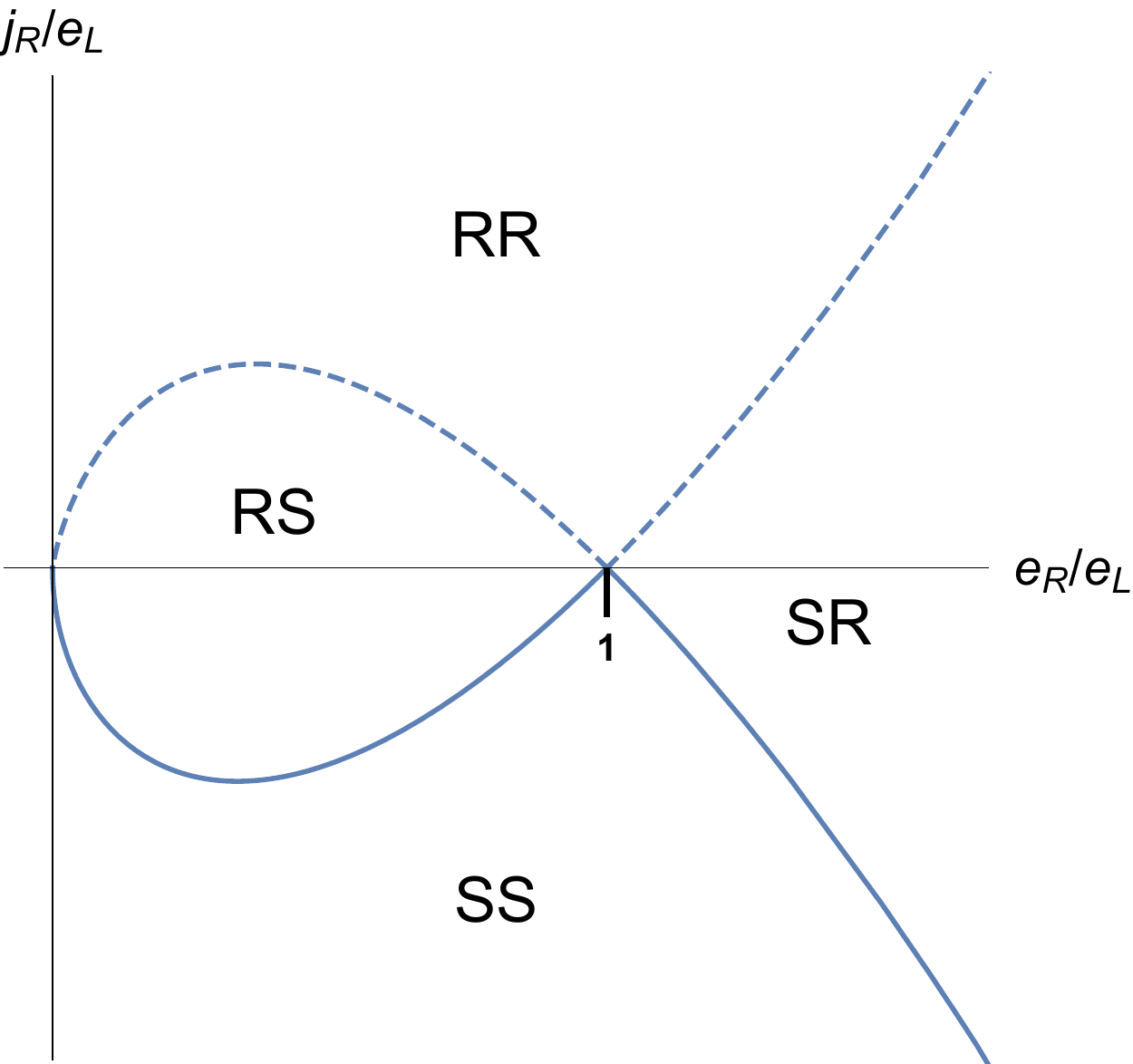}
\caption{
A phase diagram for the solution to the Riemann problem in a large $d$ limit. 
Given a pair $(e_L,0)$ and $(e_R,j_R)$,
the selection of shock and rarefaction waves is
determined by the value of $e_R/e_L$ and $j_R/e_L$. The dashed and solid lines are ``critical'':  The dashed line indicates the values of $(e_R, j_R)$ connected to $(e_L, 0)$ by a single rarefaction wave while the solid line indicates the values of $(e_R, j_R)$ connected to $(e_L,0)$ by a single shock wave. 
}
\label{fig:pd}
\end{figure}

As it happens, there are extensive treatments of this type of Riemann problem in hydrodynamics textbooks.  See for example ref.\ \cite{Smoller}.  Typically, a pair of rarefaction and/or shock waves form and move away from each other, creating in their wake a region with almost constant $e$ and $j$.  In recent literature, this intermediate region has been called a non-equilibrium steady state (NESS) \cite{Bernard:2012je, Bernard:2016nci}.  One of the main results of this paper is a ``phase'' diagram valid in a large $d$ limit (see figure \ref{fig:pd}) that describes, given the conservation equations (\ref{E:Tmn}) and initial conditions (\ref{E:ini}), which pair of waves are formed: rarefaction-shock (RS), shock-shock (SS), shock-rarefaction (SR), or rarefaction-rarefaction (RR). 
A physical reason for the preference of a rarefaction wave to a shock wave is entropy production.

Recent interest in this type of Riemann problem was spurred by a study of the problem in $1+1$ dimensional conformal field theory \cite{Bernard:2012je} where the evolution is completely determined by the conformal symmetry and a hydrodynamic limit need not be taken.  Conservation and tracelessness of the stress tensor imply that the stress tensor is a sum of right moving and left moving parts. When $j_R=j_L=0$ one finds a NESS in between the two asymptotic regions, characterized by an energy density 
$(e_R + e_L)/2$
and an energy current proportional to 
$e_R - e_L$.
The NESS is separated from the asymptotic regions by outward moving shock waves traveling at the speed of light.  (An extension of the analysis of \cite{Bernard:2012je} which includes a discontinuity in the center of mass velocity, holomorphic currents and chiral anomalies can be found in \cite{Chang:2013gba}. 
An analysis of shock waves and their relation to two dimensional turbulence was carried out in \cite{Liu:2010jg}.) 

In more than two space-time dimensions, conformal symmetry alone is not enough to specify the evolution completely and one needs additional assumptions about the structure of the conserved currents.   
Recent work appealed to the gauge/gravity duality \cite{Bhaseen:2013ypa,Amado:2015uza,Megias:2015tva,Bakas:2015hdc}, an analogy with $1+1$ dimensions \cite{Chang:2013gba}, and hydrodynamics \cite{Bhaseen:2013ypa,Pourhasan:2015bsa,Lucas:2015hnv,Spillane:2015daa}.
These papers focused on the case  $j_R=j_L=0$ and $e_L>e_R$ such that from a hydrodynamic perspective a left moving rarefaction wave and a right moving shock wave are expected to emerge.

The distinction between rarefaction and shock waves was ignored in some of these papers \cite{Bhaseen:2013ypa,Chang:2013gba,Pourhasan:2015bsa}. Indeed, when working with $2+1$ or $3+1$ dimensional conformal field theories, the difference between, say, an SS solution to the Riemann problem and an RS solution to the Riemann problem is very small for all but extreme initial energy differences. As the spacetime dimension $d$ increases however, the difference between a rarefaction wave type of solution and a shock wave solution becomes significant \cite{Spillane:2015daa}.  This amplification of the difference between the two solutions serves as a motivator for studying this Riemann problem in a large number of dimensions.

Interestingly, a large $d$ limit has independently been a topic of recent interest \cite{Emparan:2015gva,Emparan:2016sjk,Bhattacharyya:2015dva,Bhattacharyya:2015fdk,
Emparan:2013moa,Emparan:2013xia,Emparan:2013oza,Emparan:2014cia,
Emparan:2014jca,Emparan:2014aba,Emparan:2015rva,Andrade:2015hpa,Romero-Bermudez:2015bma} in the study of black hole solutions to Einstein's equations. Of particular relevance to our work is the connection between black holes in asymptotically AdS spaces and hydrodynamics \cite{Bhattacharyya:2008jc}.  Certain strongly interacting conformal field theories are known to have dual classical gravitational descriptions. In the limit where these conformal field theories admit a hydrodynamic description, a solution to the relevant hydrodynamic equations can be mapped to a solution of Einstein's equations, in a gradient expansion where physical quantities change slowly in space and time.  Transport coefficients such as shear viscosity are fixed by the form of Einstein's equations. Thus, one may study the Riemann problem in conformal field theories with a large number of dimensions by studying an equivalent Riemann-like problem involving an initially discontinuous metric of a black hole in an asymptotically AdS background.  

Given that extensive analyses of conservation equations like (\ref{E:Tmn}) can be found in many hydrodynamics textbooks and papers, one can legitimately ask why we bother to redo the analysis here.  
The reason is that when working in a large number of dimensions, one can solve for the black hole metric exactly, independent of the derivative expansion (which is naturally truncated), thus obtaining an exact solution to the Riemann problem which includes possible viscous terms and is in general valid even when gradients of thermodynamic quantities are large (as is the case with discontinuous initial conditions).

Our work is organized as follows.  In section \ref{sec:holodual}, we rederive the equations (\ref{E:Tmn}) by taking a large $d$ limit of Einstein's equations.  We show how to rewrite them as the conservation condition on a stress-tensor, $\partial_\mu T^{\mu\nu}=0$.   In section \ref{sec:hydro}, we compare the large $d$ stress tensor and equations of motion to those arising from the fluid-gravity correspondence \cite{Bhattacharyya:2008jc}. We find that both eqs.\ (\ref{E:Tmn}) and the stress tensor $T^{\mu\nu}$ are equivalent to the hydrodynamic equations that come from the fluid-gravity correspondence at large $d$, at least up to and including second order gradient corrections.  
In the same section we also construct an entropy current $J_S^\mu$ using an area element of the black hole horizon and show that the divergence of the entropy current is positive $\partial_\mu J_S^\mu \geq 0$ in this large $d$ limit.
In section \ref{sec:NESS}, we solve the Riemann problem for eqs.\ (\ref{E:Tmn}) and derive the phase diagram given in figure \ref{fig:pd}.  Finally, we conclude in section \ref{sec:discussion} with some directions for future research.
Appendix \ref{A:shockentropy} contains a short calculation of the entropy produced across a shock, while appendix \ref{A:bestiary} contains plots of auxiliary numerical results.

\section{The holographic dual of the Riemann problem for large $d$}
\label{sec:holodual}
We wish to construct a holographic dual of the Riemann problem. Consider the Einstein Hilbert action
\begin{equation}
\label{E:action}
	S = -\frac{1}{2\kappa^2} \int \sqrt{-g} \left(R + \frac{(d-2)(d-1)}{L^2} \right) \d^dx\,.
\end{equation}
A canonical stationary solution of the resulting equations of motion is the black brane solution
\begin{equation}
\label{E:sol}
	\d s^2 = 2 \d t \, \d r - r^2 \left(1 - \left(\frac{4\pi T}{(d-1) r} \right)^{d-1}\right)\d t^2 + r^2 \d x_{\bot}^2\,,
\end{equation}
where $T$ is an integration constant which denotes the Hawking temperature. The solution \eqref{E:sol} is dual to a thermal state of a conformal field theory with temperature $T$. For instance, the thermal expectation value of the stress tensor in such a state is given by 
\begin{equation}
\label{E:dual}
	\langle T^{\mu\nu} \rangle = \begin{pmatrix} (d-2) P(T) & 0 & \ldots & 0 \\ 0 & P(T) & \ldots & 0 \\ \vdots & \vdots & \ddots  & \vdots \\ 0 & 0 & \ldots & P(T) \end{pmatrix}
\end{equation}
where 
\begin{equation}
	P(T) = p_0 \left(\frac{4 \pi T}{d-1} \right)^{d-1}
\end{equation}
is the pressure with $p_0$ a theory dependent dimensionless parameter.  (The indices $\mu$ and $\nu$ run over the $d-1$ dimensions of the $(d-1)$-dimensional CFT.)

As discussed in \cite{Amado:2015uza} a dual description of the Riemann problem necessitates an initial black hole configuration which is held at some fixed temperature $T_L$ for all $z<0$ and at a different temperature $T_R$ for $z>0$. This would correspond to a configuration where the expectation value of the stress tensor is given by \eqref{E:dual} with $T=T_L$ for $z<0$ and by \eqref{E:dual} with $T=T_R$ for $z>0$.
Since the initial black hole is out of equilibrium it will evolve in time. 
 Its dual description will provide a solution for the time evolution of the stress tensor which we are after. Thus, our goal is to solve the equations of motion following from \eqref{E:action} and use them to construct the dual stress tensor. 
 
An ansatz for the metric which is compatible with the symmetries and our  initial conditions is given by 
\begin{equation}
\label{E:ansatz}
	\d s^2 =  \d t (2 \d r-g_{tt}\d t-2 g_{tz} \d z)  +  g_{zz} \d z^2 +  g_{\bot \bot} \d x_{\bot}^2\,,
\end{equation}
where the metric components are functions only of $t$, $r$, and $z$.
(A more general ansatz which involves a transverse velocity can be found in \cite{Emparan:2015gva}.)
A numerical solution of the equations of motion for $g_{tt}$, $g_{tz}$ and $g_{ii}$ ($i = x_\bot$ or $z$)
with smoothened initial conditions has been obtained for $d=4$ in \cite{Amado:2015uza} for relatively small initial temperature differences, $(T_L-T_R)/(T_L+T_R)<1$. A solution for finite $d>4$ and for large temperature differences, $(T_L-T_R)/(T_L+T_R) \sim 1$ is challenging. 

In this work we use the methods developed in \cite{Emparan:2015gva,Emparan:2016sjk} (see also 
\cite{Bhattacharyya:2015dva,Bhattacharyya:2015fdk,
Emparan:2013moa,Emparan:2013xia,Emparan:2013oza,Emparan:2014cia,
Emparan:2014jca,Emparan:2014aba,Emparan:2015rva})
 to address the Riemann problem in the limit that $d$ is very large. Such a limit can be understood as follows. In an appropriate gauge, the near boundary expansion of the metric gives
\begin{align}
\begin{split}
\label{E:nearb}
	g_{tt} &= r^2 + \mathcal{O}(r^{3-d})  \ , \\
	g_{tz} & = \mathcal{O}(r^{3-d}) \ , \\
	g_{ii} & = r^2 + \mathcal{O}(r^{3-d})\,.
\end{split}
\end{align}
Thus, in the large $d$ limit at any finite value of $r$, the spacetime looks like the AdS vacuum. Only by keeping $R = r^n$ finite with $n\equiv d-1$ will the $O(r^{-n})$ corrections to the metric remain observable. Our strategy is to solve the equations of motion in the finite $R$ region subject to the boundary conditions (\ref{E:nearb}). 
Following \cite{Emparan:2015gva}, we also use the scaling $x_{\bot} = \chi/\sqrt{n}$ and $z=\zeta/\sqrt{n}$ so that in this coordinate system the line element takes the form
\begin{equation}
	\d s^2 =  \d t (2\d r-g_{tt}\d t-2g_{t \zeta} \d \zeta)  + g_{\zeta \zeta} \d\zeta^2+ g_{\bot \bot} \d\chi_{\bot}^2  \,,
\end{equation}
where
\begin{align}
\begin{split}
\label{E:Ansatz}
	\frac{g_{tt}}{ r^2} &=  \sum_{k=0} \frac{E^{(k)}}{n^{k}} \ ,  \\
	 g_{t\zeta}  & = \sum_{k=1} \frac{J^{(k)}}{n^k} \ , \\
	 \frac{g_{ii}}{ r^2} & =  \frac{1}{n} + \sum_{k=2} \frac{g_i^{(n)}}{n^k} \,.
\end{split}
\end{align}
(In a slight abuse of notation $i$ is now either $\chi_\bot$ or $\zeta$.)
We have used the letters $E$ and $J$ to emphasize these quantities' (soon to be seen) close connection with an energy density and energy current in the dual hydrodynamic description.

One can now solve the equations of motion order by order in $1/n$.
The equations of motion are simply Einstein's equations in the presence of a negative cosmological constant:
 \begin{equation}
 \label{E:EinsteinEOM}
R_{MN} = -(d-1) g_{MN}  \ ,
\end{equation}
setting $L=1$ for convenience.
Let $a$ and $b$ index the $t$, $r$, and $\zeta$ directions only, while $i$ and $j$ index the remaining perpendicular directions.
Furthermore, let $\tilde R_{ab}$ be the Ricci tensor with respect to the three dimensional metric in the $t$, $r$, and $\zeta$ directions.
Then
\begin{eqnarray}
R_{ab} &=& \tilde R_{ab} + \frac{d-3}{4} (\partial_a \log g_{\bot \bot}) (\partial_b \log g_{\bot \bot}) - \frac{d-3}{2} \frac{\nabla_a \partial_b g_{\bot \bot}}{g_{\bot \bot}} \ , \\
R_{ij} &=& \delta_{ij} \left( \frac{5-d}{4} \frac{(\partial_a g_{\bot \bot})(\partial^a g_{\bot \bot})}{g_{\bot \bot}} - \frac{1}{2} \nabla^a \partial_a g_{\bot \bot} \right) \ .
\end{eqnarray}

Imposing that the boundary metric is Minkowski and choosing a near boundary expansion of the form \eqref{E:nearb} we find
\begin{align}
\begin{split}
\frac{g_{tt}}{r^2} &=
1 - \frac{e}{R}
- \frac{1}{n} \left( \frac{e_2}{R} + \frac{\log R}{R} \partial_\zeta j + \frac{j^2}{2R^2} \right) +O(n^{-2}) \ , \\
g_{t\zeta} &= \frac{1}{n} \frac{j}{R} + \frac{1}{n^2} \left( 
\frac{j_2}{R} 
 + \frac{\log R}{R} \left( \partial_\zeta \left( \frac{j^2}{e} \right) +2 f \right)
+ \frac{j^3}{2 R^2 e}
\right) + O(n^{-3})\ , \\
\frac{g_{\zeta \zeta}}{r^2} &= \frac{1}{n} + \frac{1}{n^2} \frac{j^2}{R e} + O(n^{-3}) 
\ , \\
\frac{g_{\bot \bot}}{r^2} &= \frac{1}{n} -  \frac{1}{n^3} \frac{j^2}{R e} + O(n^{-4}) \ ,
\end{split}
\end{align}
where the $\mathcal{O}(n^{-2})$ correction to $g_{tt}$ and the $\mathcal{O}(n^{-3})$ contributions to $g_{\zeta \zeta}$ are too long to write explicitly. The functions $e$ and $j$ are functions of $t$ and $\zeta$ only and must satisfy the additional constraints
(\ref{E:Tmn}).
Equations \eqref{E:Tmn} are identical to those obtained in \cite{Emparan:2015gva,Emparan:2016sjk}. We can rewrite them in terms of a conservation law
\begin{equation}
	\partial_{\mu}T^{\mu\nu} = 0
\end{equation}
where
\begin{equation}
\label{E:fullTmn}
T^{\mu\nu} = \begin{pmatrix}
		e & j - \partial_\zeta e \\
		j - \partial_\zeta e & \quad e + \frac{j^2}{e} - 2 \partial_\zeta j + \partial_\zeta^2 e
		\end{pmatrix} + \begin{pmatrix} \partial_\zeta^2 g & - \partial_\zeta \partial_t g \\ - \partial_\zeta \partial_t g & \partial_t^2 g \end{pmatrix} \,.
\end{equation}
where $g$ is an arbitrary function.
Likewise, the functions $e_2$ and $j_2$ must also satisfy a set of equations which can be obtained from the conservation of
\begin{multline}
\label{E:fullT2mn}
	T_2^{\mu\nu} = 
\begin{pmatrix}
 e_2 &
  \left(\frac{j^2}{e} +e -e_2 - 2j' \right)'
  +j+j_2 \\
  \left(\frac{j^2}{e}  + e - e_2 - 2j' \right)'
   +j+j_2&
	T^{11}
	\end{pmatrix} 
	\\
	+ \begin{pmatrix} \partial_\zeta^2 g_2 & - \partial_\zeta \partial_t g_2 \\ - \partial_\zeta \partial_t g_2 & \partial_t^2 g_2 \end{pmatrix} \,.
\end{multline}
where
\begin{multline}
	T^{11} =   2 \left( 1- \left(\frac{j}{e} \right)' \right) \left( \frac{j}{e} \right)' e (\log(e) -3)
	+e + e_2 \left( 1- \frac{j^2}{e^2} \right)+ 2 j_2 \frac{j}{e} \\
	+ \left( e_2- 4 e - 6 \frac{j^2}{e}  + 4 j' \right)''  
	- 2 \left(j_2 -  3 j - \frac{j^3}{e^2} \right)' + \frac{j^2}{e} \left( \frac{j}{e} \right)' 
 \ . 
 \end{multline}
 We will use $'$ and $\partial_\zeta$ interchangeably in what follows.

\section{Comparison with hydrodynamics}
\label{sec:hydro}

Let us pause to understand \eqref{E:fullTmn}. Within the context of the gauge-gravity duality it is possible to construct a solution to the Einstein equations which is perturbative in $t$, $\zeta$ and $\chi_{\bot}$ derivatives of the metric components \cite{Bhattacharyya:2008jc}.  Such a perturbative solution to the equations of motion, which is available for any dimension $d$ \cite{Haack:2008cp,Bhattacharyya:2008mz}, allows for a dual description of the theory in terms of fluid dynamical degrees of freedom. 

\subsection{Stress tensor from fluid-gravity correspondence}
\label{SS:Stresstensor}
To construct the dual hydrodynamic description of a slowly varying black hole, we boost the black hole solution \eqref{E:sol} by a constant velocity $u^{\mu}$ in the $t$, $z$, $x_{\bot}$ directions. The resulting line element is given by
\begin{equation}
\label{E:LEO0}
	\d s_{(0)}^2 = 2 u_{\mu}\d x^{\mu} \d r - r^2 \left(1 - \left(\frac{4 \pi T}{(d-1)r}\right)^{d-1} \right)u_{\mu}u_{\nu}\d x^{\mu}\d x^{\nu} + r^2 \left(\eta_{\mu\nu} + u_{\mu}u_{\nu} \right)\d x^{\mu}\d x^{\nu}\,.
\end{equation}
Allowing for $u^{\mu}$ and $T$ to become spacetime dependent implies that \eqref{E:LEO0} will get corrected. By setting gradients of $u^{\mu}$ and $T$ to to be small, one can solve for the corrections to \eqref{E:LEO0} order by order in derivatives so that the line element will take the schematic form
\begin{equation}
	\d s^2 = \d s_{(0)}^2 + \d s_{(1)}^2 + \ldots
\end{equation}
where $\d s_{(i)}^2$ denotes the $i$th order gradient corrections to the line element. 

The stress tensor $T^{\mu\nu}$ which is dual to \eqref{E:LEO0} takes the form
\begin{equation}
\label{E:Tmnhydro}
	T^{\mu\nu}  = \sum_i T_{(i)}^{\mu\nu} 
\end{equation}
also expanded in gradients. One finds \cite{Haack:2008cp,Bhattacharyya:2008mz}
\begin{equation}
	T^{\mu\nu}_{(0)} = P(T) \left( (d-1) u^{\mu} u^{\nu} + \eta^{\mu\nu} \right)
\end{equation}
which is nothing but a boosted version of \eqref{E:dual} and then, in the Landau frame,
\begin{align}
\begin{split}
	T^{\mu\nu}_{(1)} &= -2 \eta \sigma_{\mu\nu} \ , \\
	T^{\mu\nu}_{(2)} & = 
	 \frac{(d-1) \eta}{2 \pi T}  \left[
	 (1-\tau_0)
	 u\cdot \mathcal{D} \sigma^{\mu\nu} + {\sigma^{\lambda}}_{\mu}\sigma_{\lambda\nu} - \frac{\sigma^{\alpha\beta} \sigma_{\alpha\beta}}{d-2} P_{\mu\nu}  
	 - \tau_{0} \left( \omega_{\mu}{}^{\lambda}\sigma_{\lambda\nu} + \omega_{\nu}{}^{\lambda}\sigma_{\mu\lambda} \right)
	 \right]
\end{split}	
\end{align}
with
\begin{align}
\begin{split}
	P^{\mu\nu} &= \eta^{\mu\nu} + u^{\mu}u^{\nu} \ , \\
	\sigma_{\mu\nu} &= \frac{1}{2} P_{\mu}{}^{\alpha}P_{\nu}{}^{\beta} \left(\partial_{\alpha}u_{\beta} + \partial_{\beta}u_{\alpha} \right) - \frac{1}{d-2} P^{\mu\nu} \partial_{\alpha}u^{\alpha}\ , \\
	\omega_{\mu\nu} & = \frac{1}{2} P^{\mu\alpha}P^{\nu\beta} \left(\partial_{\alpha}u_{\beta} - \partial_{\beta}u_{\alpha} \right)\ ,  \\
	u\cdot \mathcal{D} \sigma_{\mu\nu} & = P_{\mu}{}^{\alpha} P_{\nu}{}^{\beta} u^{\lambda} \partial_{\lambda} \sigma_{\alpha\beta} + \frac{\partial_{\alpha}u^{\alpha}}{d-2} \sigma_{\mu\nu}\ , 
\end{split}
\end{align}
and
\begin{equation}
\label{transportcoeffs}
	\eta = \frac{(d-1)P}{4 \pi T} \  ,
	\qquad
	\tau_0 = \int_1^{\infty} \frac{y^{d-3} -1}{y(y^{d-1}-1)} dy = \frac{1}{2} + O(d^{-2})\,.
\end{equation}	
(Note that our definition of $\sigma_{\mu\nu}$ is somewhat unconventional.) An initial analysis of third order gradient corrections has been carried out in \cite{Grozdanov:2015kqa} for $d=5$. A full analysis of all third order transport terms for arbitrary dimension $d$ is currently unavailable.

Since \eqref{E:fullTmn} has been obtained from a large $d$ limit of a gravitational dual theory, we expect that \eqref{E:fullTmn} coincides with \eqref{E:Tmnhydro} when the former is expanded in derivatives and the latter is expanded around large $n=d-1$. 
In short, we expect that taking a gradient expansion commutes with taking a large $d$ limit.
To make a direct comparison let us consider the hydrodynamic stress tensor \eqref{E:Tmnhydro} in the $t$, $\zeta$, $\chi_\bot$ coordinate system where the metric tensor takes the form
\begin{equation}
	ds^2 = -\d t^2 + \frac{\d \zeta^2}{n} + \frac{\d \chi_{\bot}^2}{n} \ .
\end{equation}
One important effect of this rescaling is to keep the sound speed to be an order one quantity.

 Scaling the spatial component of the velocity field by $1/\sqrt{n}$, viz.,
\begin{equation}
	u^{\mu} = \frac{1}{\sqrt{1-\frac{\beta^2(t,\zeta)}{n}}} \left( 1 ,\,\beta(t,\zeta)\right)\,,
\end{equation}
and maintaining that $\epsilon = (d-2)P$ is finite in the large $d$ limit, we find,
\begin{align}
\begin{split}
	\sigma^{\mu\nu} &=n \partial_\zeta \beta \,\delta^{\mu}_{\zeta} \delta^{\nu}{}_{\zeta} +\mathcal{O}(n^0)\\
	u \cdot \mathcal{D} \sigma^{\mu\nu} &= n \left(\beta \partial_{\zeta}^2 \beta + \partial_{t}\partial_{\zeta} \beta\right)  \,\delta^{\mu}_{\zeta} \delta^{\nu}{}_{\zeta} +\mathcal{O}(n^0)\\
	{\sigma^{\lambda}}_{\mu}\sigma_{\lambda\nu} - \frac{\sigma^{\alpha\beta} \sigma_{\alpha\beta}}{d-2} P_{\mu\nu} & = n \left(\partial_{\zeta} \beta \right)^2 \,\delta^{\mu}_{\zeta} \delta^{\nu}{}_{\zeta} +\mathcal{O}(n^0)
\end{split}
\end{align}
and thus,
\begin{equation}
\label{E:TmnO2}
	T^{\mu\nu} = \begin{pmatrix}  \epsilon & \beta \epsilon \\ \beta \epsilon &\quad  \epsilon (1+ \beta^2 ) + p \end{pmatrix} + \mathcal{O}\left(n^{-1}\right)
\end{equation}
where
\begin{equation}
	p =  - 2 \epsilon \partial_{\zeta} \beta + 2 \epsilon (\partial_{\zeta}\beta)^2 + \epsilon \beta \partial_{\zeta^2}\beta+ \epsilon \partial_{\zeta} \partial_t \beta + \mathcal{O}\left(\partial^3\right)
\end{equation}
and $\mathcal{O}\left(\partial^3\right)$ denotes third order and higher derivative corrections. 
Note that this constitutive relation for the stress tensor includes and encodes the large $d$ limit of the transport coefficients
(\ref{transportcoeffs}).  

Now, we insert the redefinitions
\begin{align}
\begin{split}
\label{E:hydroexpansion}
	e & = \epsilon - \frac{1}{2} \partial_\zeta^2 \epsilon \  , \\
	j & = \beta \epsilon + \partial_{\zeta}\epsilon + \frac{1}{2} \partial_t \partial_\zeta \epsilon \ , \\
	g & = \frac{1}{2} \epsilon
\end{split}
\end{align}
into the large $d$ constitutive relation for the stress tensor \eqref{E:fullTmn}, use the large $d$ stress tensor conservation equations (\ref{E:Tmn}), and throw out terms that have three or more derivatives.  We claim that in this fashion, we recover the stress tensor \eqref{E:TmnO2} in the gradient expansion. Thus, the large $d$ limit and the gradient expansion seem to commute. Note that while the conservation equations \eqref{E:Tmn} are of second order in gradients of $\zeta$ and $t$, the stress tensor includes at least second order gradients. 

The implications of  \eqref{E:hydroexpansion} are worth emphasizing. The equations of motion \eqref{E:Tmn} are equivalent to the standard equations of motion of relativistic hydrodynamics when the latter are expanded in a large $d$ limit. When working with the $e$ and $j$ variables one obtains equations of motion which are second order in derivatives and therefore include dissipative effects. When carrying out a frame transformation to the more traditional Landau frame, more derivatives will appear. When considering the stress tensor associated with the equations of motion \eqref{E:Tmn} one obtains more terms with higher gradients which do not contribute to the equations of motion. It would be interesting to see if one can construct an alternative to the Israel-Stewart theory using a ``large $d$-frame'' where gradients naturally truncate.

\subsection{Entropy from Gravity}
\label{SS:Entropy}
Within the context of our forthcoming analysis, it is instructive to compute the dual entropy production rate which is associated with the evolution of the horizon. Due to its teleological nature, it is usually difficult to identify the location of the event horizon. However, in the large $d$ limit the analysis is somewhat simplified.
Let us look for a null surface of the form $R=r_h(t,\zeta)$. The normal to such a surface is
\begin{equation}
	\Xi_M \d x^{M} = \d R - \partial_t r_h \d t - \partial_\zeta r_h \d\zeta \ .
\end{equation}
Demanding that $\Xi^2 \Big|_{R=r_h} = 0$ implies, to leading order in the large $d$ limit, that
\begin{equation}
\label{E:EH1}
	r_h = e \,.
\end{equation}
The spacetime singularity which exists in our solution implies that an event horizon must be present. Since the only null surface available is \eqref{E:EH1}, it must be the location of the event horizon.
Subleading corrections to the location of the event horizon are given by
\begin{align}
\begin{split}
	r_h &= e + \frac{1}{n} \left(
	\frac{4 j e' - 2 (e')^2 - j^2}{2 e} + e_2 - 2 j' + 2 e'' + j' \log(e)	
	\right)
	\\
   & \equiv e + \frac{1}{n} r_{h\,1} \ .
\end{split}
\end{align}

To compute the change in the black hole entropy over time we compute the area form of the event horizon.  Following the prescription of \cite{Bhattacharyya:2008xc}, we find that
\begin{equation}
	\mathbf{A} = \frac{\epsilon_{\mu_1 \ldots \mu_d}}{(d-1)!} J_S^{\mu_1} \d x^{\mu_2} \wedge \ldots \wedge \d x^{\mu_{d}}
\end{equation}
where
\begin{equation}
	J_S^{\mu} =\frac{\sqrt{h}}{4 G_N} \frac{N^{\mu}}{N^t}
\end{equation}
where $h$ is the spatial $(t=\hbox{constant})$ part of the induced metric on the horizon
\begin{equation}
	H_{\mu\nu}\d x^\mu \d x^\nu = g_{MN}\d x^{M}\d x^{N}\Big|_{R=r_h}
\end{equation}
and $N^\mu$ is defined via
\begin{equation}
	\Xi^M \partial_M = N^R \partial_R + N^\mu \partial_\mu\,.
\end{equation}
A short computation yields
\begin{align}
\begin{split}
	\sqrt{h} &=n^{-\frac{n-1}{2}} \left(e + \frac{1}{n} \left(r_{h\,1} -e \ln e  \right) \right) \ , \\
	N_\mu \d x^\mu &= -\partial_t e \, \d t - \partial_{\zeta} e \, \d\zeta \ .
\end{split}
\end{align}
Thus,
\begin{multline}
\label{E:JSgravity}
	\tilde{J}_S^{\mu} = 16\pi G_N n^{\frac{n-1}{2}} J_S^{\mu} = \frac{4\pi}{n} \begin{pmatrix} e,\, & j  - e',\, & \ldots \end{pmatrix}
		 \\
		+\frac{4\pi}{n^2} \begin{pmatrix} r_{h\, 1} - e \ln e  ,\, & 
		\left(\frac{j^2}{2e^2} + \log e \right) (2 e'-j) +  \left(\frac{j^2}{e} \right)'  \log e+ j_2 - r_{h \, 1}'
		,\, & \ldots \end{pmatrix}
\end{multline}
where we have normalized the entropy density so that it is compatible with our conventions for the energy density. 

The second law of black hole thermodynamics amounts to
\begin{equation}
	\partial_{\mu}J_S^{\mu} \geq 0 \ .
\end{equation}
In our large $d$ limit we find that
\begin{equation}
\label{E:divJgravity}
	\partial_{\mu} \tilde{J}_S^\mu =  \frac{8 \pi e}{n^2} 
	\left[ \partial_\zeta \left( \frac{j - \partial_\zeta e}{e} \right) \right]^2 \, .
\end{equation}
The expectation from hydrodynamics, to second order in derivatives, is that the divergence of the entropy current is given by
\begin{equation}
\label{E:Jsrule}
	\partial_\mu \tilde{J}_S^\mu = \frac{2\eta}{T}\sigma^2\,.
\end{equation}
(See for example (8) of ref.\  \cite{Son:2009tf}.)
This expectation matches \eqref{E:divJgravity} on the nose. Note that to leading order in the large $d$ limit the entropy current vanishes. This somewhat surprising feature of the large $d$ limit follows from the fact that entropy production terms are suppressed by inverse powers of the dimension in the large $d$ limit.
Another way of understanding this suppression comes from thinking about the temperature $T \sim e^{1/(d-1)}$.  In the large $d$ limit, $T$ is constant to leading order in $d$.  From the thermodynamic relation $\d e = T \d s$, it then follows that changes in energy are proportional to changes in entropy, and entropy conservation follows from energy conservation at leading order in a large $d$ expansion.\footnote{%
We thank R.~Emparan for a discussion on this point.
}

\section{Near equilibrium steady states}
\label{sec:NESS}

We now analyze the dynamics controlled by the partial differential equations (\ref{E:Tmn})
which encode the dynamics of an out of equilibrium black hole \eqref{E:ansatz} and its dual stress tensor \eqref{E:fullTmn}. Various related holographic analyses can be found in \cite{Chesler:2009cy,Chesler:2010bi,Chesler:2013lia,Balasubramanian:2013yqa,Chesler:2015fpa,Khlebnikov:2010yt,Shuryak:2012sf,Fischetti:2012vt,Figueras:2012rb,Emparan:2013fha}.
As discussed in the introduction, the particular question we would like to address is a Riemann problem:  What is the time evolution following from an initial condition \eqref{E:ini}?
We are particularly interested in the steady state solution which will emerge at late times. For convenience we will consider a reference frame for which $j_L=0$.
Indeed, if $e(x,t)$ and $j(x,t)$ satisfy the conservation equations (\ref{E:Tmn}), then so do $e(x-v t,t )$ and $j(x- v t, t) + v e(x - v t, t)$. Thus, for constant values of $e$ and $j$, we can choose a $v$ such that $j$ will be set to zero. The non-relativistic nature of the boost symmetry reflects the fact that the large $d$ limit we have taken is effectively a non-relativistic limit where the speed of light $c \sim \sqrt{d}$ has been pushed off to infinity.

\subsection{Rarefaction waves vs.\ shock waves}
\label{SS:rareshock}
\begin{figure}[htb]
\centering
\includegraphics[width=2.0in]{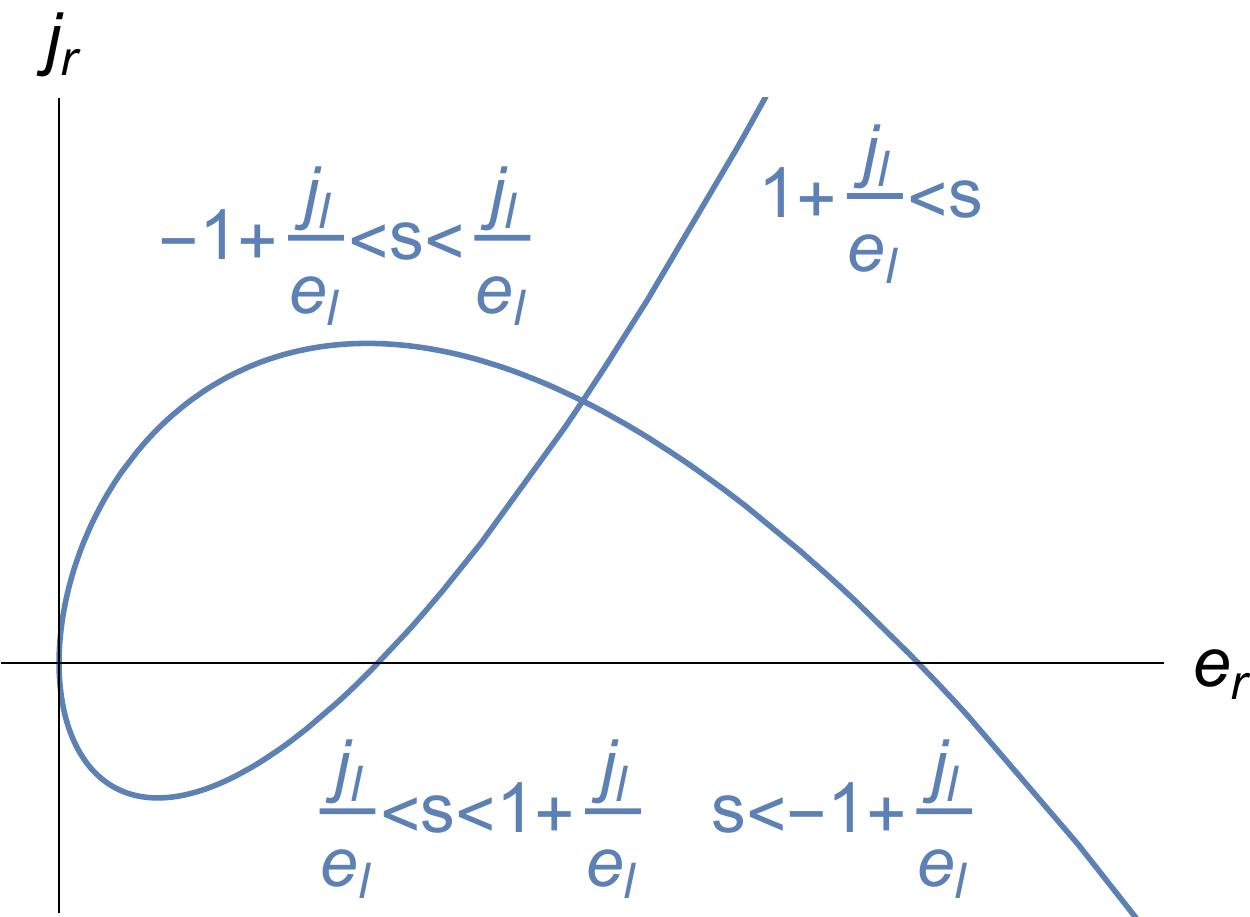}
\hfill
\includegraphics[width=2.0in]{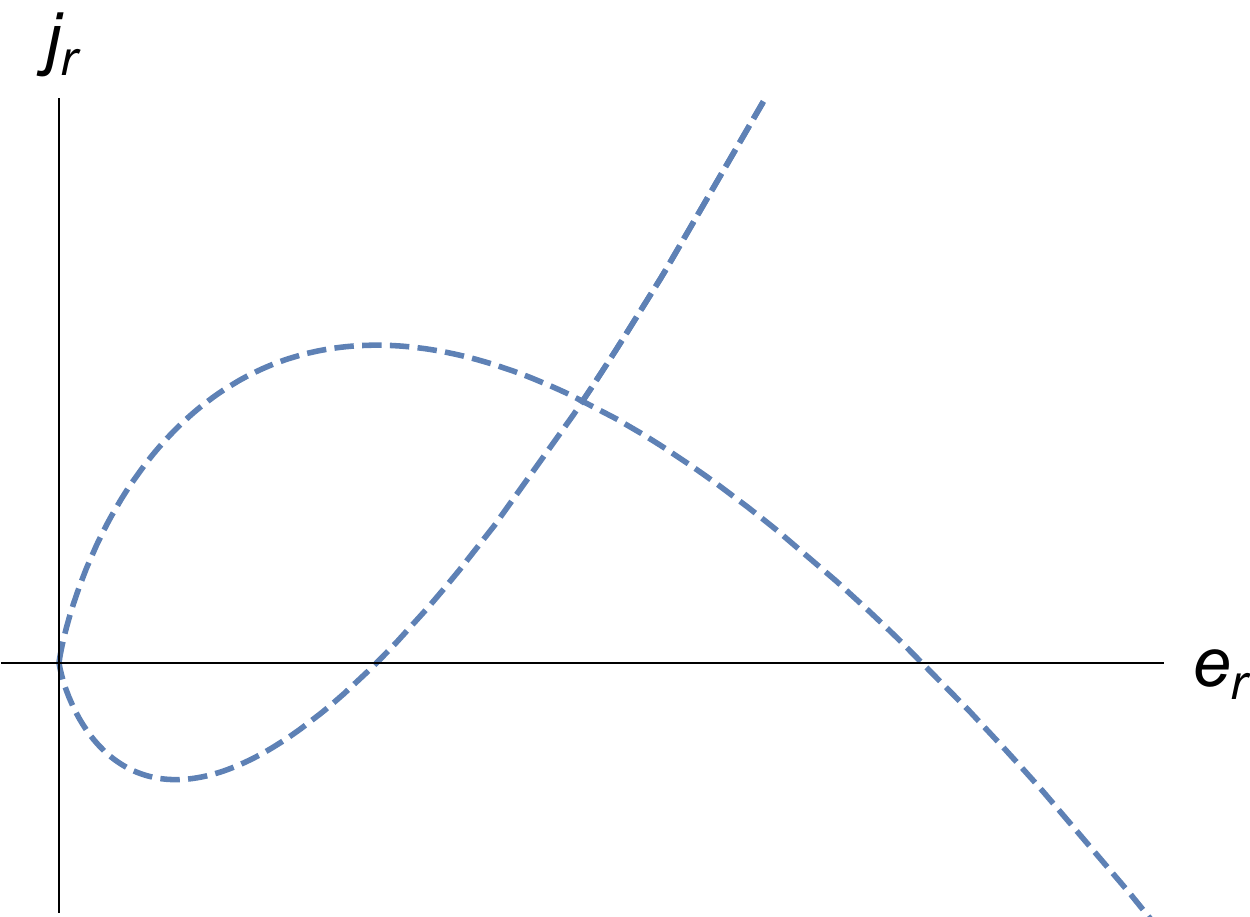}
\hfill
\includegraphics[width=2.0in]{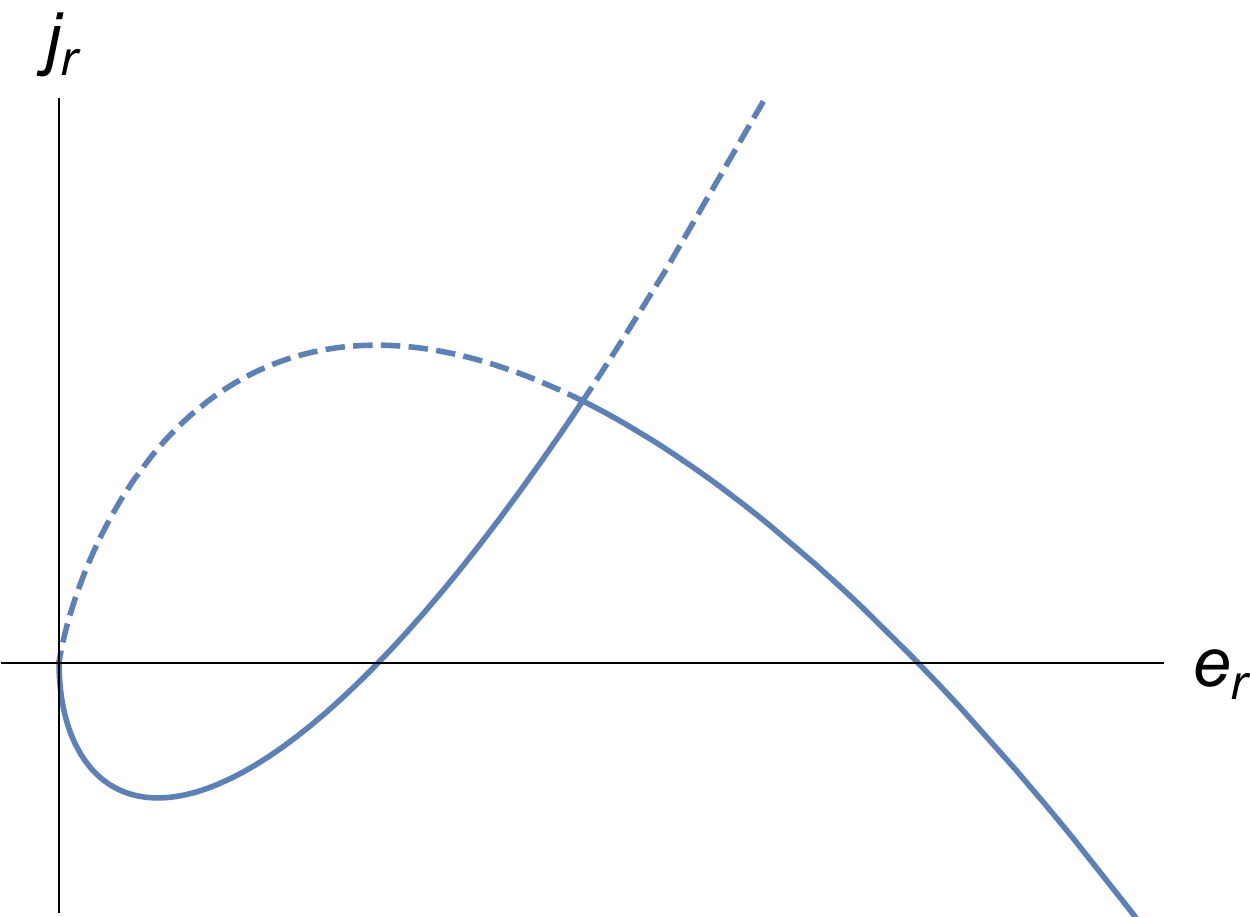}
\caption{
(Left panel) The solid blue curve corresponds to the solution to the Rankine-Hugoniot condition for $(e_l,j_l)$. Points on the curve correspond to different values of $s$ in \eqref{E:ourRH}. The regions $j_l/e_l < s < j_l/e_l +1$ and $s<j_l/e_l-1$ correspond to good shocks satisfying \eqref{E:goodshock1} and \eqref{E:goodshock2} respectively. (Center panel) The dashed line, which almost overlaps with the blue line at places, parameterizes the rarefaction solution \eqref{E:rarefaction} also associated with $(e_l,j_l)$. (Right panel) A plot of possible values of $(e_r,j_r)$ for a given a pair $(e_l,j_l)$ with good shocks preferred over the rarefaction solution and the rarefaction solution preferred over bad shocks.}
\label{fig:simplefish}
\end{figure}

Before addressing the Riemann problem in its entirety let us consider a simplified system which is less constrained. Consider \eqref{E:fullTmn} with gradient terms neglected. The resulting expression is the large $d$ limit of the energy momentum tensor of an inviscid fluid which is known to support (discontinuous) shock waves \cite{Smoller} for any finite value of $d$.  
While the solution to the full Riemann problem will consist of a pair of shock and/or rarefaction waves, we begin in this section with a single discontinuous shock wave moving with velocity $s$. Conservation of energy and momentum imply
\begin{equation}
\label{E:RHeqns}
	s [ T^{tt} ] = [ T^{t\zeta} ] \ ,
	\qquad
	s [ T^{t\zeta} ] = [ T^{\zeta\zeta} ] \ ,
\end{equation}
where $\left[ Q \right] = Q_l - Q_r$ and $Q_{r/l}$ specify the value of $Q$ to the left or right of the shock respectively.\footnote{%
 In this section we use subscripts $r$ and $l$ to denote values of quantities to the right or left of the shock. In other 
 sections we use subscripts $R$ and $L$ to denote quantities in the right and left asymptotic regions. In the latter case there is generally an 
 interpolating region which we denote with a $0$ subscript.
}
The conservation conditions \eqref{E:RHeqns} are very general and are often referred to as the Rankine-Hugoniot (RH) relations. In our setup they reduce to
\begin{align}
\begin{split}
\label{E:ourRH}
	s e_l - j_l &= s e_r - j_r  \ ,\\
	s j_l - \left( e_l + \frac{j_l^2}{e_l} \right) &= s j_r - \left( e_r + \frac{j_r^2}{e_r} \right) \ ,
\end{split}
\end{align}
where $e_{r/l}$ and $j_{r/l}$ are the energy density and current immediately to the right or left of the shock.
While these Rankine-Hugoniot relations hold for an arbitrary, piece-wise continuous fluid profile, in what follows, we are interested in the much simpler situation where $e$ and $j$ are constant functions away from the shocks.
Amusingly, $e_r$ satisfies a cubic equation,\footnote{%
 In general $d$, one finds the relation
 \[
 \sinh^2(\alpha_l - \alpha_r) = \frac{d-2}{(d-1)^2} \frac{(\epsilon_l - \epsilon_r)^2}{\epsilon_l \epsilon_r} \ ,
 \]
 where $\beta = \tanh \alpha$ is the fluid velocity. }
\begin{equation}
\label{E:cubic}
	(e_l j_r - e_r j_l)^2 = e_l e_r (e_l- e_r)^2 \ ,
\end{equation}
a plot of which as a function of $j_r$ resembles a fish: fixing $(e_l,j_l)$, each value of $s$ is mapped to a point on the $(e_r,j_r)$ plane. The collection of such points is given by a fish-like curve, an example of which is given in the left panel of figure \ref{fig:simplefish}.

We make two observations about the fish.
The vacuum $(e_r, j_r) = (0,0)$ always lies on the cubic \eqref{E:cubic}, corresponding to the fact that a shock can interpolate between
any value of $(e_l, j_l)$ and the vacuum.  
Also $(e_r, j_r) =(e_l, j_l)$ is the point of self-intersection of the cubic and has 
$s=\pm 1+j_l/e_l$.  The physical content of this observation is that when $(e_r, j_r)$ is close to $(e_l, j_l)$ but still lies on the cubic, we can find a close approximation to the fluid profile by linearizing the equations of motion.  As we will describe in greater detail below, linearized fluctuations correspond to damped sound modes, and indeed the two regions can be connected by sound waves propagating at the local sound speed $s = \pm 1 + j_l/e_l$.  

The shock solutions we found all solve the conservation equations \eqref{E:ourRH}. However, some of these solutions are unphysical in the following sense. Let us boost to a frame where the shock speed vanishes, $s=0$. In half of the shock solutions, a quickly moving fluid at low temperature is moving into a more slowly moving fluid at higher temperature, converting kinetic energy into heat and producing entropy.  We will refer to these shocks as ``good'' shocks. The other half of the solutions correspond to the time reversed process where a slowly moving fluid at high temperature moves into a rapidly moving but cooler fluid, turning heat into kinetic energy.  This second solution, as we shall see shortly, should be discarded.

Strictly speaking, entropy is conserved in the large $d$ limit (see the discussion following equation \eqref{E:Jsrule}). A more formal way of understanding why one should discard the bad shocks is to restore the gradient corrections but take a limit where these are small. 
Let us assume that in the frame where the shock velocity is zero there is an approximately stationary configuration such that time derivatives are much smaller than spatial derivatives. Boosting back to a shock with velocity $s$, we expect that $e$ and $j$ depend only on the combination $\zeta-s t$, i.e., $j(t,\zeta) = j(\zeta-s t)$ and likewise, $e(t,\zeta) = e(\zeta-s t)$.
The equations of motion \eqref{E:Tmn} become ordinary differential equations which can be integrated once to obtain
\begin{align}
\begin{split}
\label{E:integratedonce}
	e' =& -s (e-e_l) + (j-j_l) \ , \\
	j'  =& - s(j-j_l) + \left( e + \frac{j^2}{e} - e_l - \frac{j_l^2}{e_l} \right) \ .
\end{split}
\end{align}
We have picked the two integration constants such that $e'$ and $j'$ vanish in the left asymptotic region.
The Rankine-Hugoniot conditions \eqref{E:ourRH} imply that $e'$ and $j'$ also vanish in the right asymptotic region.
As $e'$ and $j'$ themselves vanish in the left and right asymptotic regions, we can describe $e'$ and $j'$ well near these points by looking at a gradient expansion.  Near the left asymptotic region
\begin{align}
\begin{split}
\label{E:stability}
\left(
\begin{array}{c}
e' \\
j' 
\end{array}
\right) 
&\approx
\left( 
\begin{array}{cc}
-s & 1 \\
1 - \frac{j_l^2}{e_l^2} &  \frac{2j_l}{e_l}-s 
\end{array}
\right)
\left(
\begin{array}{c}
e-e_l \\
j - j_l 
\end{array}
\right)\ \\
&\equiv M_l
\left(
\begin{array}{c}
e-e_l \\
j - j_l 
\end{array}
\right)\,.
\end{split}
\end{align}
There is a similar looking equation for $e'$ and $j'$ near the right asymptotic region
\begin{equation}
\label{E:stability2}
\left(
\begin{array}{c}
e' \\
j' 
\end{array}
\right) 
\approx
M_r
\left(
\begin{array}{c}
e-e_r \\
j - j_r 
\end{array}
\right)\,. 
\end{equation}

The solutions near $(e_l, j_l)$ and near $(e_r,j_r)$ have an exponential nature with the sign of the exponents depending on the eigenvalues of  $M_l$ and $M_r$ appearing on the right hand side of \eqref{E:stability} and \eqref{E:stability2} given by 
\begin{equation}
\label{E:eigenvalues}
	\lambda_{r\,\pm} = \pm 1 +\frac{j_r}{e_r} -s  \ , \qquad \lambda_{l\,\pm} = \pm 1 + \frac{j_l}{e_l} - s\,.
\end{equation} 
We now observe that the signs of the eigenvalues of $M_l$ and $M_r$ determine whether the shock is a viable solution to the equations of motion.
\begin{itemize}
\item
If both eigenvalues of $M_l$ are negative, then $e'$ and $j'$ will not vanish as $x \to -\infty$. Thus we require that at least one eigenvalue of $M_l$ is positive in order for a shock solution to exist.
\item
If we assume there is exactly one positive eigenvalue, then $1+j_l/e_l >s$ and $-1 + j_l/e_l < s$.
Note that the value $1+j_l/e_l$ corresponds to the slope of one of the characteristics (i.e.\ the local speed of one of the sound waves), and this condition implies that this characteristic will end on the shock.  Since $\lambda_{l\,-}$ is assumed to be negative, we have to tune one of the two integration constants of the system of differential equations to zero.  This tuning means that generically the solution to the right of the shock will be a linear combination of both of the solutions near $(e_r, j_r)$.  If both solutions are to be used, then it had better be that both eigenvalues of $M_r$  are negative.  (Otherwise, it will not be true that $e'$ and $j'$ vanish in the limit $x \to \infty$.)  In particular, the larger of the two eigenvalues must be negative, which implies that $1+j_r/e_r < s$. (In terms of characteristics, both will end on the shock.) Thus, we find the constraint
\begin{subequations}
\label{E:goodshocks}
\begin{equation}
\label{E:goodshock1}
	1+j_r/e_r < s < 1 + j_l/e_l\,.
\end{equation}
\item
If both eigenvalues of $M_l$ are positive, we still need at least one negative eigenvalue of $M_r$ to be able to connect the solutions in the left and right asymptotic regions. Moreover, for $M_r$ to have two negative eigenvalues would be inconsistent with momentum conservation \eqref{E:ourRH}. An analysis similar to the previous one yields
\begin{equation}
\label{E:goodshock2}
	-1+j_r/e_r < s < -1 + j_l/e_l\,.
\end{equation}
\end{subequations}
\end{itemize}
The constraints \eqref{E:goodshocks} choose the good shocks over the bad ones.\footnote{%
 In appendix \ref{A:shockentropy}, we discuss a third RH relation one can write down for the entropy current.  
 If the RH relations for energy and momentum are satisfied, the RH relation for the entropy current will typically be violated due to entropy production %, albeit at subleading order in a $1/d$ expansion
associated with viscous effects.  In the weak shock limit, we demonstrate that gradient corrections produce the entropy that leads to this violation of the third RH relation. 
 Reversing the sign  of the energy difference between the two asymptotic regions in eqs.\ (\ref{sprodstationary}) or (\ref{Eproduction}), it is straighforward to see that a bad shock would lead to a decrease in entropy, at least in the simple case where $s = 0$ and $j_r = j_l$.
 }

Since bad shocks are not allowed, one may inquire as to the time evolution of a discontinuity with initial conditions which would have generated a bad shock. As it turns out, bad shocks can be replaced by the more physical rarefaction solutions \cite{Smoller}. The rarefaction solution assumes that between the asymptotic regions specified by $(e_l,j_l)$ and $(e_r,j_r)$, there is an interpolating solution where $e$ and $j$ are functions of $\xi= \zeta/t$.  As was the case for the shock wave, given $e_l$ and $j_l$, there is a one parameter family of allowed values of $e_r$ and $j_r$. These are given by 
\begin{align}
\begin{split}
\label{E:rarefaction}
	e_r =& e_l \exp \left( \pm j_l/e_l - 1 \mp \xi_r \right) \ , \\
	j_r =& e_l ( \pm 1+\xi_r) \exp \left(\pm j_l/e_l - 1 \mp \xi_r \right) \ .
\end{split}
\end{align}
The curve traced by $(e_r, j_r)$ also resembles a fish, and 
for moderate values of the shock parameters $e_r$ and $j_r$ it closely follows the cubic curve corresponding to a shock solution.  (See the central panel of figure \ref{fig:simplefish}.)  
The vacuum $(0,0)=(e_r,j_r)$ solution can always be connected to  $(e_l, j_l)$ through a rarefaction wave.
The self-intersection point $(e_r,j_r) = (e_l, j_l)$ has $\xi = \mp 1 + j_l/e_l$, again corresponding to a sound wave type interpolation between the two regions $(e_r, j_r) \approx (e_l, j_l)$.  

Given that bad shocks are replaced by rarefaction waves, one should remove from the fish diagram (left panel of figure \ref{fig:simplefish}) the portion of the curve which corresponds to bad shocks and replace it with a curve corresponding to a rarefaction solution (central panel of figure \ref{fig:simplefish}). The resulting curve can be found on the right panel of figure \ref{fig:simplefish}: the belly of the fish and the lower part of its tail corresponds to a good shock and its back and upper tail to a rarefaction solution. One may compute the curve explicitly by imposing \eqref{E:goodshocks}, but it can also be understood from a graphical viewpoint as we now explain.

Recall that the self intersection point of the shock wave fish (solid curve on the left panel of figure \ref{fig:simplefish}) corresponds to a shock velocity, $s$, which takes the values of the local speed of sound, $\pm 1 + j_l/e_l$.  On the tail, $s$ is either larger than $1 + j_l/e_l$ (upper tail) or smaller than $-1 + j/e$ (lower tail).  Thus, on the tails, the eigenvalues are either both positive or both negative. The top portion of the tail has $\lambda_{\pm l} < 0$ while the bottom portion of the tail has $\lambda_{\pm l} > 0$.  As a result, the top portion of the tail must be replaced by a rarefaction wave while the bottom portion can be a shock.  To decide which portion of the body of the shock fish to replace by a rarefaction wave, one must study $\lambda_{\pm r}$.  

Consider a second fish which exhibits the solution to the cubic \eqref{E:cubic} for a given value of $(e_r, j_r)$.  We will call this second fish an $r$-fish and the first an $l$-fish. Similar to the analysis of the tail of the $l$-fish, we find that the bottom portion of the tail of the $r$-fish should be constructed from a rarefaction solution while the top portion from a shock.

\begin{figure}
\centering
\includegraphics[width=2.5in]{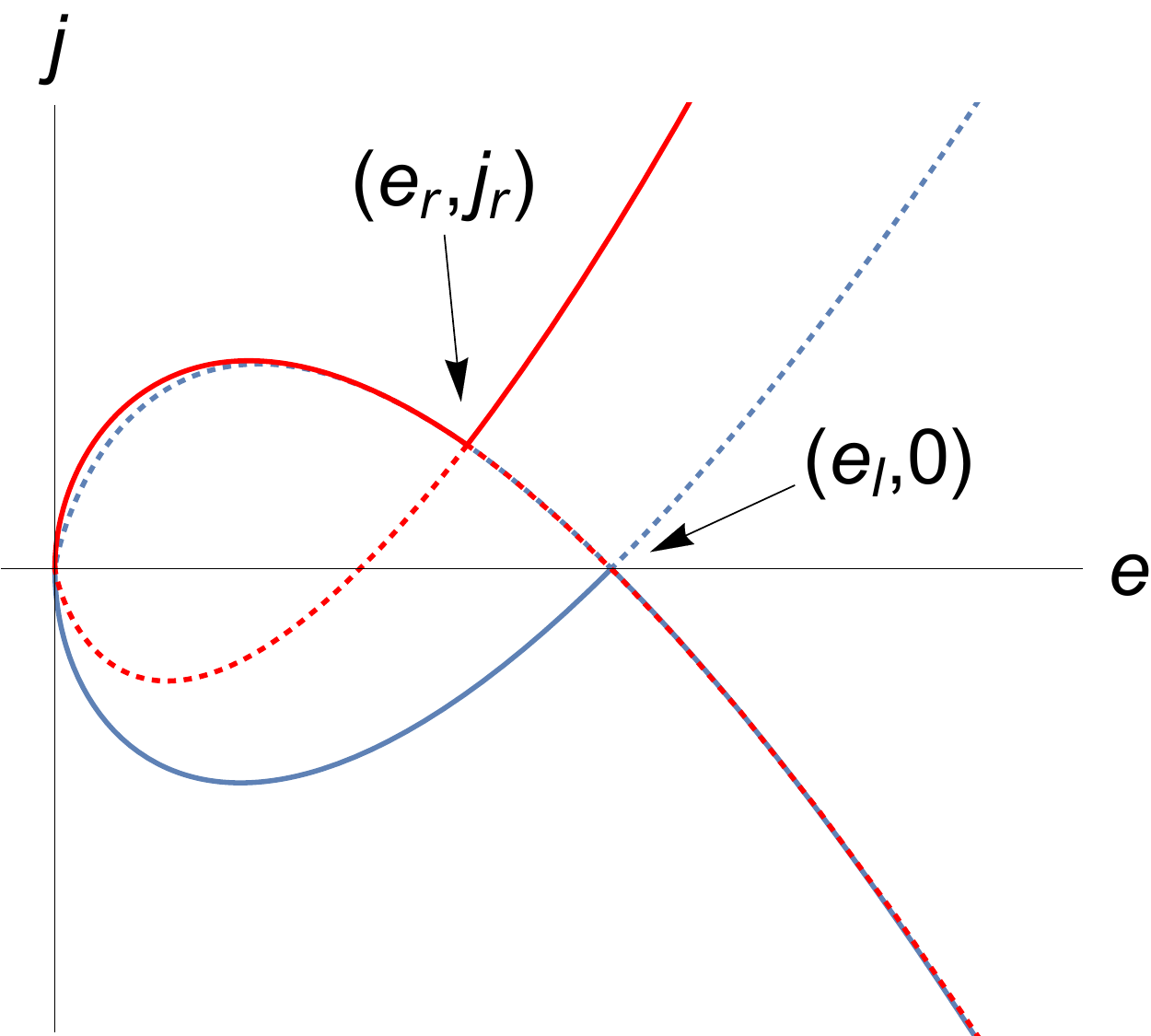}
\phantom{bbbbb}
\includegraphics[width=2.5in]{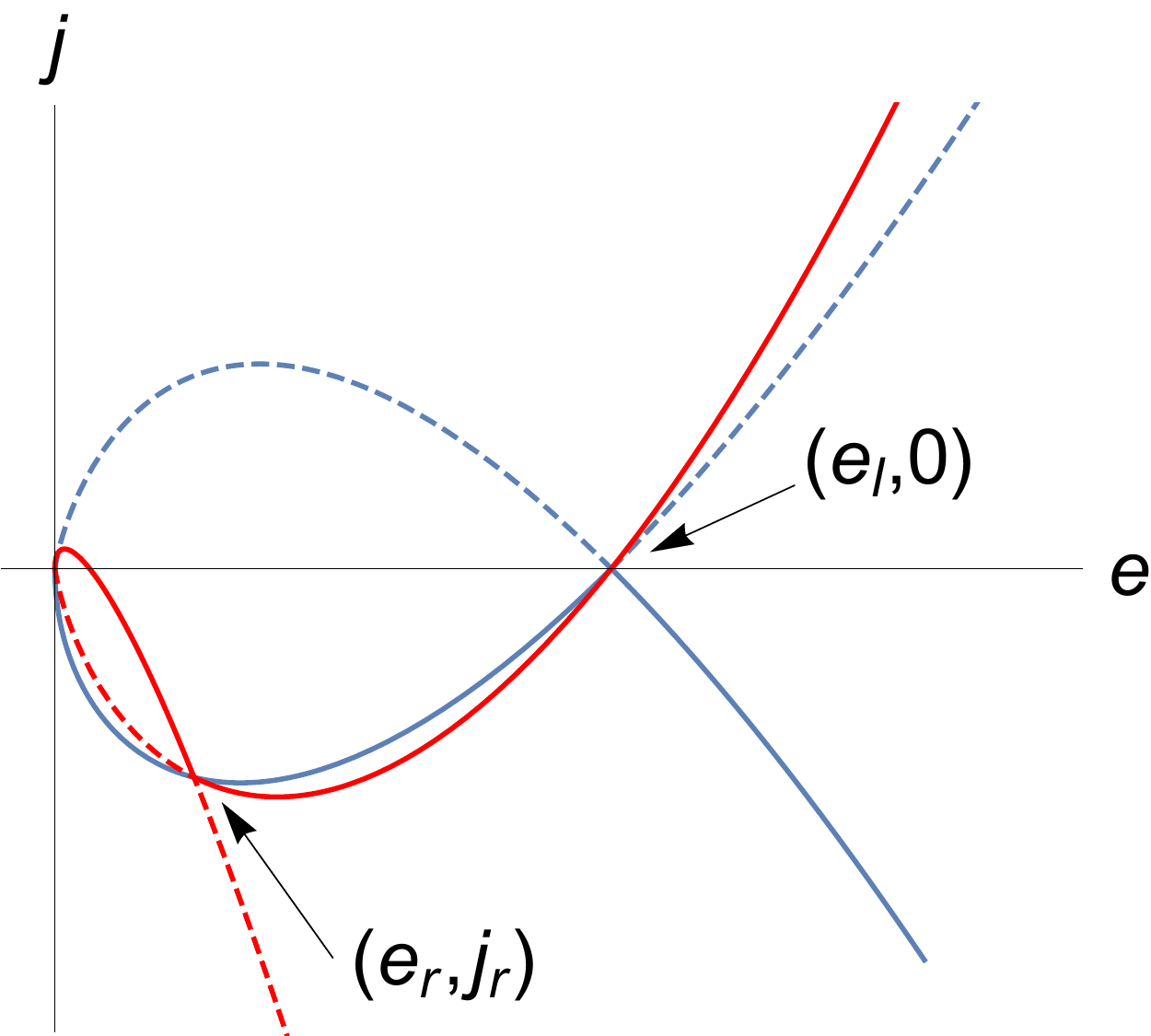}
\caption{A graphical determination of the ``good shocks'' and ``bad shocks''.  The red fish corresponds to $(e_r, j_r)$ while the blue fish is built from $(e_l, 0)$. See the main text for a discussion.
}
\label{fig:si}
\end{figure}

Consider an $r$-fish whose point of self intersection lies somewhere on the body of the $l$-fish. When the $r$-fish is drawn so that it intersects the back of the $l$-fish, the bottom portion of the $r$-fish's tail will go through the point of self-intersection of the $l$-fish (see the left panel of figure \ref{fig:si}).  As the bottom portion of the tail of the $r$-fish is a rarefaction, the region $(e_r, l_r)$ can be connected to $(e_l, j_l)$ by a rarefaction.  Reciprocally, since we're describing a single shock or rarefaction interface between two regions, the back of the $l$-fish should be replaced by a rarefaction wave. We can run the argument again for an $r$-fish drawn to intersect the belly of the $l$-fish.  We conclude that the belly of the $l$-fish must be a shock (see the right panel of figure \ref{fig:si}). 

\subsection{Solving the Riemann problem using ideal hydrodynamics}
\label{SS:implicit}

Armed with our understanding of shock waves and rarefaction solutions, let us now tackle the Riemann problem we set out to solve. At $t=0$, we consider a pair $(e_L,0)$ which describes the fluid for $z < 0 $ and another pair $(e_R, j_R)$ describing the fluid for $z >0$.  For a single interpolating shock or rarefaction, we have seen that given $(e_L,0)$ there is a one parameter family of solutions that determine $(e_R,j_R)$. Thus, generically, there will not be a single shock or rarefaction solution that joins $(e_L, 0)$ to an arbitrary $(e_R, j_R)$. However, we can connect the two regions using a pair of shock and/or rarefaction waves. That is, we could connect $(e_L,0)$ to an intermediate regime with values of $e$ and $j$ given by $(e_0,j_0)$ using a shock or rarefaction wave and another shock wave or rarefaction wave to connect the intermediate regime to the right asymptotic region $(e_R,j_R)$. In all cases, given the initial conditions, the pair of rarefaction and/or shock waves should be such that they move away from each other. 

The strategy for determining which type of solution is allowed is to prefer good shocks over rarefaction solutions and rarefaction solutions over bad shocks. Thus, given a pair $(e_L,0)$ and $(e_R,j_R)$ we need to establish which of the four possibilities for the time evolution of the initial state is allowed: 
two shocks (SS), a rarefaction wave followed by a shock (RS),
or the remaining two configurations which we will denote by SR and RR. 

\begin{figure}
\centering
\includegraphics[width=1.8in]{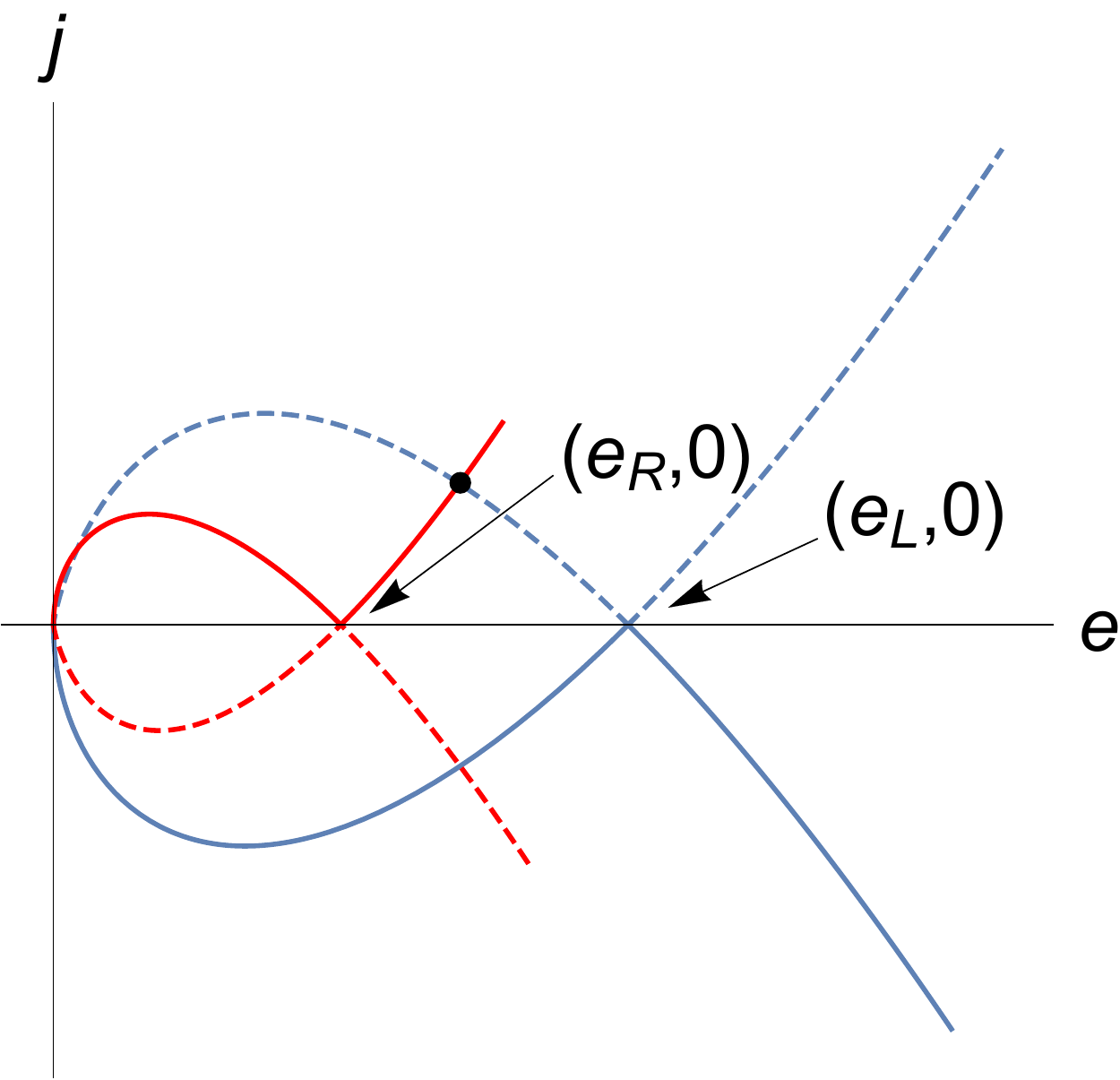}
\hfill
\includegraphics[width=1.8in]{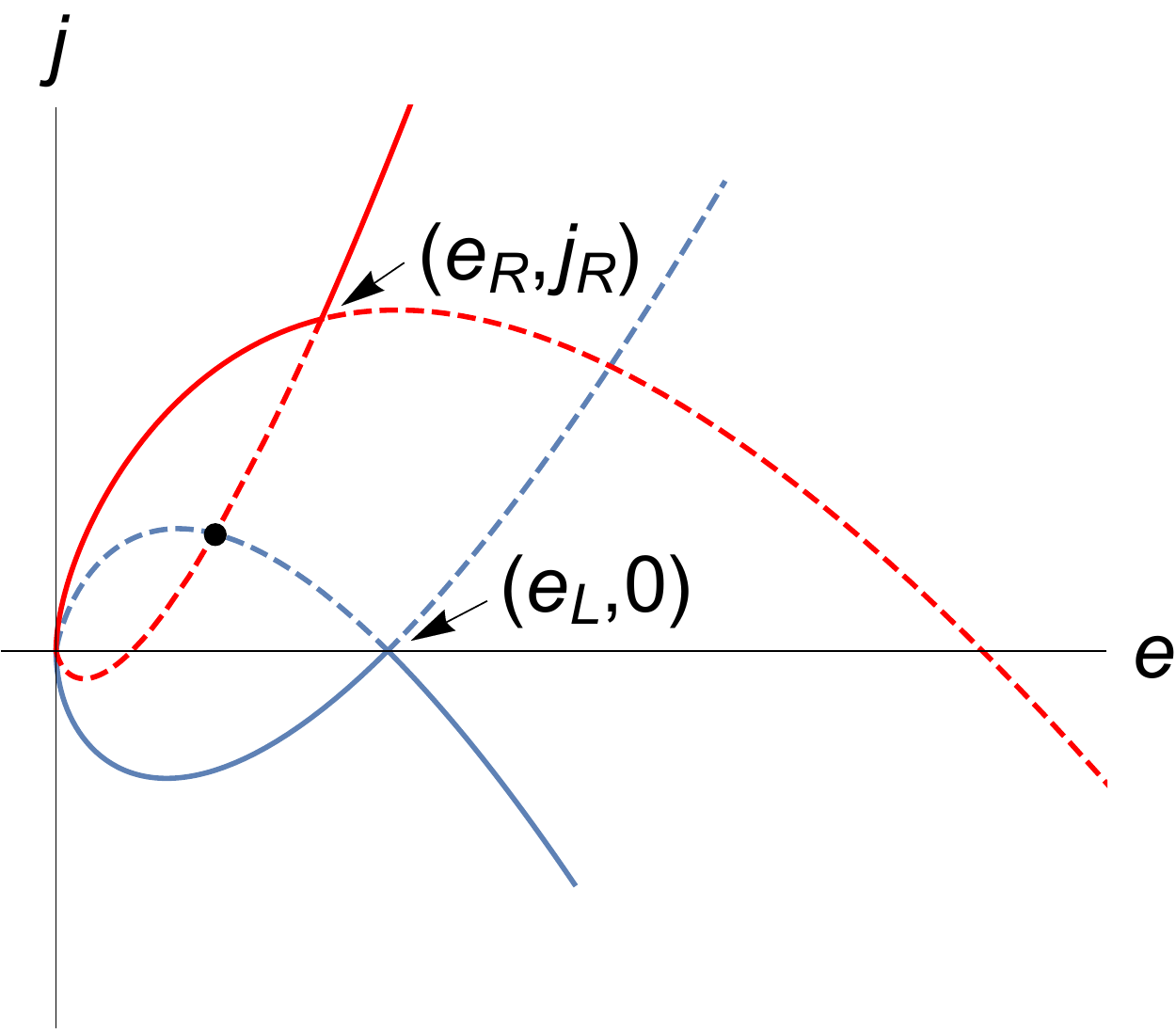}
\hfill
\includegraphics[width=1.8in]{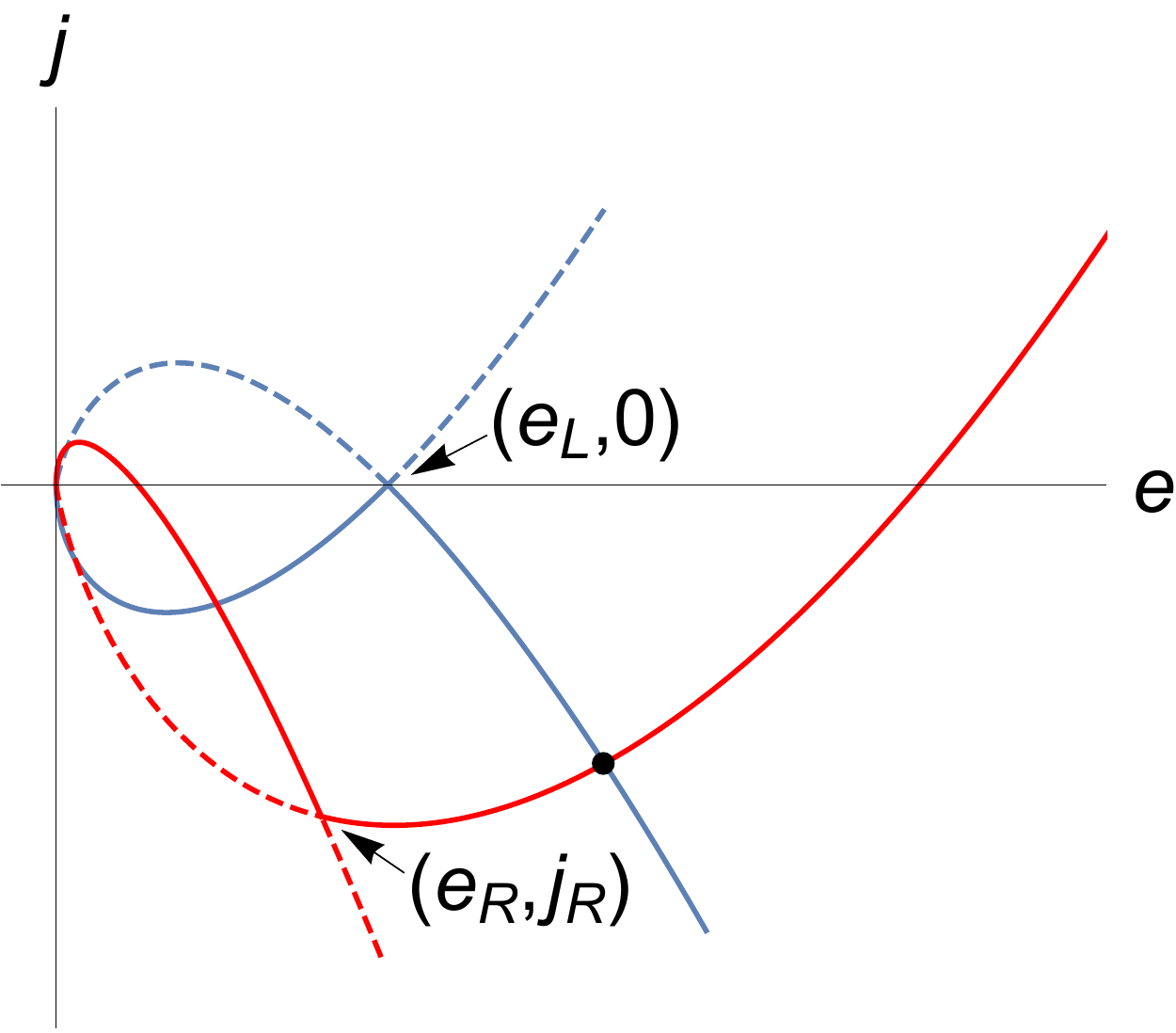}
\caption{Some diagramatic solutions to the Riemann problem.  The blue fish corresponds to $(e_L,0)$ while the red fish to $(e_R, j_R)$.  The solid line is a shock and the dashed line a rarefaction.  The intermediate region is indicated by a black dot. Left panel: The shock solution of the right asymptotic region overlaps with the rarefaction solution of the left asymptotic region, so we get an SR type configuration. Center panel: The rarefaction solution of the left and right regions overlap creating an RR type solution. Right panel: An SS type solution.
}
\label{fig:eg}
\end{figure}

To understand the possible solutions to the Riemann problem, let us first consider two fish diagrams: one associated with $(e_l,j_l) = (e_L,0)$ (the $l$-fish) and another with $(e_r,j_r) = (e_R,j_R)$ (the $r$-fish). The points of overlap of the diagrams will give us the possible value of $e_0$ and $j_0$. We will always choose a point where the two disturbances are moving away from each other. See, for example, figure \ref{fig:eg}. 

Instead of plotting the $r$- and $l$-fishes, we can obtain closed form expressions for the various types of solutions by solving \eqref{E:goodshocks} and \eqref{E:rarefaction} on a case by case basis. In the following we provide some simple examples of such expressions.
\begin{itemize}
\item{{\bf RS configurations.}}
As an example of the RS case, we take $(e_L,0)$ and $(e_R,0)$ as the asymptotic regions with $e_L>e_R$. The SR case is a left-right reflection of the RS case and therefore does not warrant further discussion.

To estimate the values of $e_0$ and $j_0$ we can follow the strategy laid out in \cite{Lucas:2015hnv,Spillane:2015daa}. For the left region we use the solution \eqref{E:rarefaction} with $e_l=e_L$, $j_l=0$, $e_r=e_0$ and $j_r=j_0$. For the right region we use \eqref{E:ourRH} with $e_l=e_0$, $j_l=j_0$, $e_r=e_R$ and $j_r=0$. We find
\begin{align}
\begin{split}
\label{E:steadystate}
	e_0 &= e_R s^2 \ ,\\
	j_0 &= e_R s (s^2 - 1) \ , \\
	0&=\frac{1}{s}-s-\log\left(\frac{e_R}{e_L}s^2 \right) \,,
\end{split}
\end{align}
which, unsurprisingly, coincides with the large $d$ limit of the hydrodynamic analysis of \cite{Lucas:2015hnv,Spillane:2015daa}. 

As pointed out in \cite{Lucas:2015hnv} the rarefaction solution will cover the location of the original shock discontinuity whenever 
\begin{equation}
 \frac{e_L}{e_R} \geq \left( \frac{1+\sqrt{5}}{2} \right)^2 \exp(1)  \sim 7.11655 \ .
\end{equation}
At the point $\zeta=0$ in the rarefaction wave, the values of $e$ and $j$ are time independent
(since any function of $\zeta/t$ will have a fixed point at $\zeta=0$). Moreover for a conserved stress tensor $T^{\mu\nu} = T^{\mu\nu}\left(\frac{\zeta}{t}\right)$, the first spatial derivative of $T^{t\zeta}$ and the first and second spatial derivatives of $T^{\zeta\zeta}$ vanish at this fixed point. Thus, one may think of the pressure at the fixed point as a ``short'' steady state for long enough times. ``Short'' implies that the region is of small spatial extent. From this perspective one has split steady states for large enough initial temperature differences. The values of $e$ and $j$ at the short steady state are given by
\begin{equation}
	e_s = j_s = e_L \exp(-1)\,.
\end{equation}

\item{{\bf SS configurations.}}
A simple example of the SS case has $(e_L,0)$ on the left and $(e_L,j_R)$ on the right with $j_R<0$.  We compute the NESS by gluing two shock waves to an intermediate region with $(e,j)=(e_0,j_0)$, similar to the RS case. Setting $\beta = j_R / e_L$, the intermediate NESS is given by 
\begin{equation}
\label{E:steadystateSS}
	e_0 = \frac{e_L}{8} (8 + \beta^2 - \beta \sqrt{16+\beta^2}) \ , \quad 
	\frac{j_0}{e_0} = \frac{\beta}{2} \ , 
\end{equation}
and the shock velocities for the left and right moving shocks, $s_L$ and $s_R$ respectively, are given by
\begin{eqnarray}
s_L &=& \frac{1}{4}(\beta - \sqrt{16+\beta^2}) \ , \\
s_R &=& \frac{1}{4}(3 \beta + \sqrt{16+\beta^2}) \ .
\end{eqnarray}

\item{{\bf RR configurations.}}
Using $e_L=e_R$ and $j_R>0$, we can find simple solutions that involve two rarefaction waves.\footnote{%
As it turns out in the RR phase, there is a simple expression for the steady state for all values of $e_L$, $e_R$, $j_L$ and $j_R$,
\begin{eqnarray*}
	e_0 = \sqrt{e_L e_R} \exp\left( \frac{j_L}{2 e_L} - \frac{j_R}{2 e_R} \right) \ , \; \; \;
	j_0 = \frac{e_0}{2} \left( \xi_+ + \xi_- \right) \,,
\end{eqnarray*}
where
\begin{eqnarray*}
	\xi_+ - \xi_- = 2\,,
	\qquad
	\xi_+ + \xi_- =\frac{j_L}{e_L} + \frac{j_R}{e_R} - \log \frac{e_R}{e_L}  \ .
\end{eqnarray*}  
A fixed point associated with a left moving rarefaction solution occurs whenever
\begin{eqnarray*}
	\frac{e_R}{e_L} \leq \exp\left( \frac{j_L}{e_L} + \frac{j_R}{e_R} -2 \right) \; \; \;
	\mbox{with} \; \; \; e_s = j_s = e_L \exp\left(-1+ \frac{j_L}{e_L}\right)  \ ,
\end{eqnarray*}
and a fixed point associated with the right moving rarefaction solution occurs whenever
\begin{equation*}
	\frac{e_R}{e_L} \geq \exp\left(\frac{j_L}{e_L} + \frac{j_R}{e_R} + 2\right)\; \; \;
	\mbox{with} \; \; \; e_s = -j_s = e_R \exp\left(-1+ \frac{j_R}{e_R} \right)\,.
\end{equation*}
}
  In this case, the NESS 
is characterized by
\begin{equation}
\label{E:steadystateRR}
	e_0 = e_L \exp\left(- \frac{j_R}{2 e_L} \right) \ , \qquad
	j_0 = \frac{j_R}{2} \exp \left( - \frac{j_R}{2 e_L} \right) \,,
\end{equation}
where the left moving rarefaction wave extends from $\xi=-1$ to $\xi = \xi_-$ while the right moving rarefaction wave extends from $\xi = \xi_+$ to $\xi = 1$ with
\begin{equation}
	\xi_+ - \xi_- = 2\,,
	\qquad
	\xi_+ + \xi_- =\frac{j_R}{e_R} \,,
\end{equation}  
Similar to the RS case we find that there is a fixed point associated with the left moving wave whenever
\begin{equation}
	\frac{j_R}{2 e_L} \geq 1\,,
\end{equation}
with
\begin{equation}
	e_s = j_s = e_L \exp(-1)\,.
\end{equation}

\end{itemize}

We claim that given $(e_L,0)$, the ``phase diagram'' of figure \ref{fig:pd} immediately allows us to choose the correct configuration of shocks and rarefaction waves for any $(e_R, j_R)$. Indeed, following figure \ref{fig:eg},  the location of the self intersection point of the $r$-fish will determine the nature of the intersection of the $r$- and $l$-fish: if the intersection point of the $r$-fish lies above the $l$-fish we will always get an RR solution; if the intersection point of the $r$-fish is below the $l$-fish we get an SS solution; and RS and SR solutions will correspond to an intersection point of the $r$-fish in the body or tail of the $l$-fish respectively. Conformal invariance dictates that the phase diagram can depend on the only two dimensionless parameters of this problem, and we obtain the phase diagram in figure \ref{fig:pd}.

Note that even though the $r$-fish and the $l$-fish intersect at $(0,0)$, we can always rule out an intermediate point that corresponds to a vacuum.  The vacuum intersection point is always along the bodies of the two fish where we have $\lambda_{-,l/r} < 0 < \lambda_{+,l/r}$.  As discussed, we can not in general connect the two asymptotic solutions if we do not have two eigenvalues of the same sign (positive for $l$ and negative for $r$) in one of the regions. 

\subsection{A numerical solution to the Riemann problem.}
\label{SS:numerical}

In the previous sections we have obtained predictions for the evolution of $e$ and $j$ starting from an initial configuration \eqref{E:ini} and assuming that gradient corrections to the equations of motion are small. It is somewhat unfortunate that this assumption stands in stark contrast to the discontinuous jump in the initial state and one may inquire whether the analysis of the previous section is relevant for the problem at hand. 
In order to resolve this issue we solve the full equations of motion \eqref{E:Tmn} numerically.  We give numerical examples of the RR, SS, and RS phases described above.  To our numerical accuracy, the difference in $e_0$ and $j_0$ between the ideal case which we have studied analytically and the case with gradients included which has been obtained numerically appears to disappear in the long time limit.

As it turns out, the equations (\ref{E:Tmn}) are easy to evolve numerically with canned PDE solvers, such as Mathematica's NDSolve routine \cite{Mathematica}.  To obtain various solutions one can evolve the initial condition  
\begin{align}
\label{E:iniRS}
	e &= \langle e \rangle \left( 1 + \delta e \tanh (c \sin (2 \pi x / L) ) \right) \, \ , \\
	j &= \langle j \rangle \left( 1 + \delta j \tanh (c \sin (2 \pi x / L) ) \right) \ , \nonumber
\end{align}
in a periodic box of length $L$.  
(In appendix \ref{A:bestiary}, we use a more elaborate piecewise continuous initial condition.)
For $c$ sufficiently large, the initial condition approaches a square wave.  As long as the disturbance has not travelled a distance of order $L$, causality ensures that the behaviour of $e$ and $j$ are very close to that of an infinite system where the values of $e$ and $j$ in the asymptotic region are fixed at some constant value. If we denote these asymptotic values as $e_L$ and $e_R$ then
\begin{equation}
	\delta e = \frac{e_L-e_R}{e_L+e_R} \; \; \; \mbox{and} \; \; \; \langle e \rangle = \frac{1}{2} \left(e_L + e_R \right)\,.
\end{equation}
We can similarly define $\langle j \rangle$ and $\delta j$.  
\begin{figure}
%\centering
\includegraphics[width=3in]{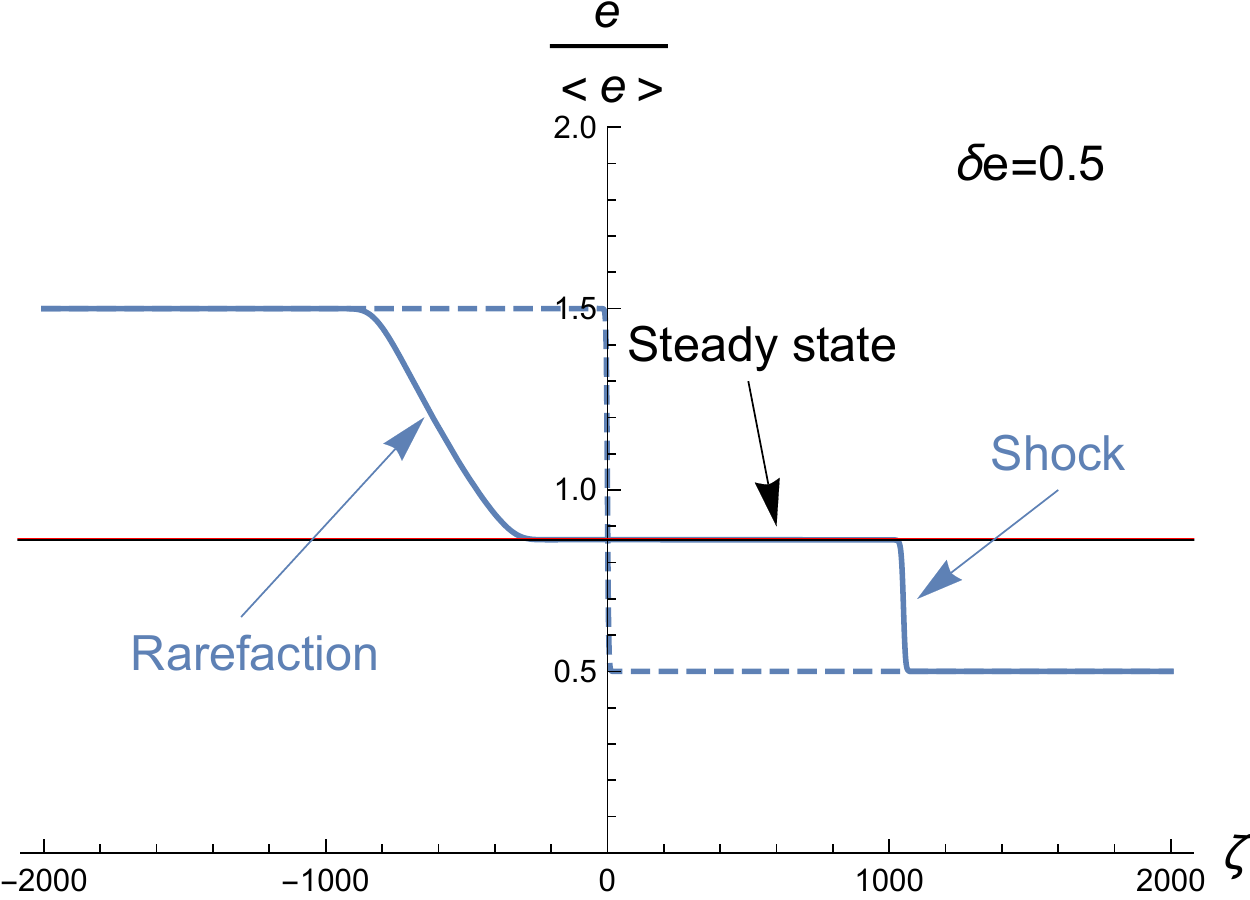}
\hfill
\includegraphics[width=3in]{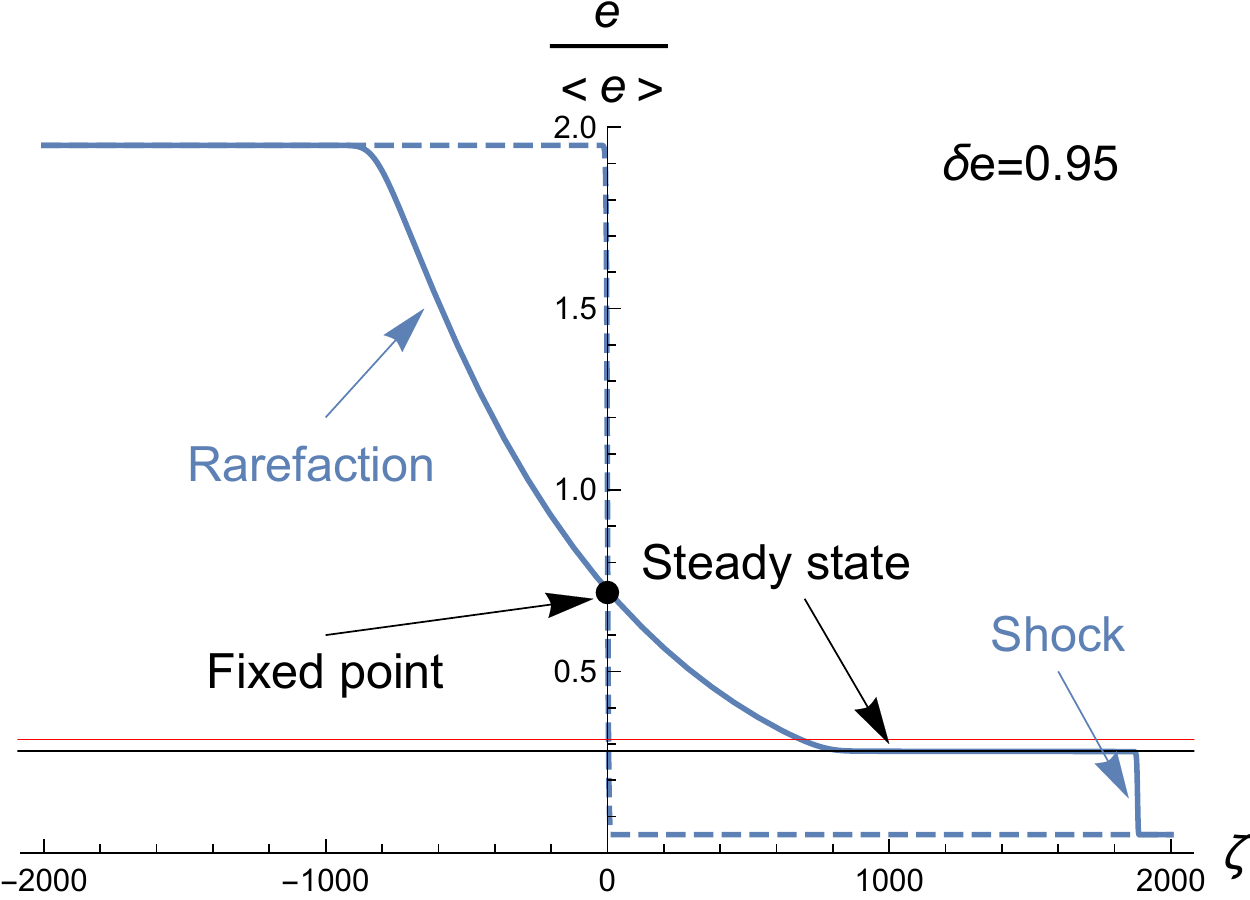}
\\
\includegraphics[width=3in]{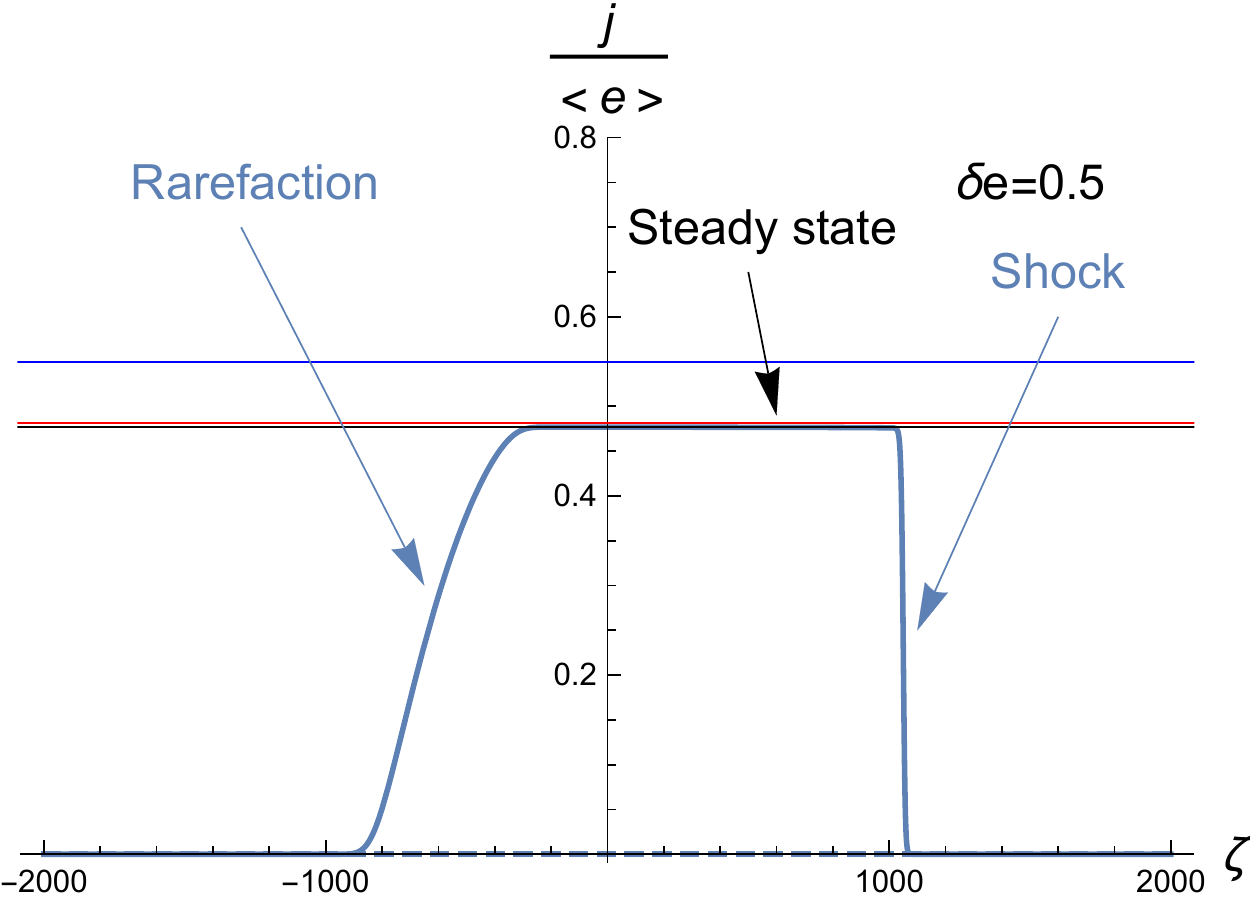}
\hfill
\includegraphics[width=3in]{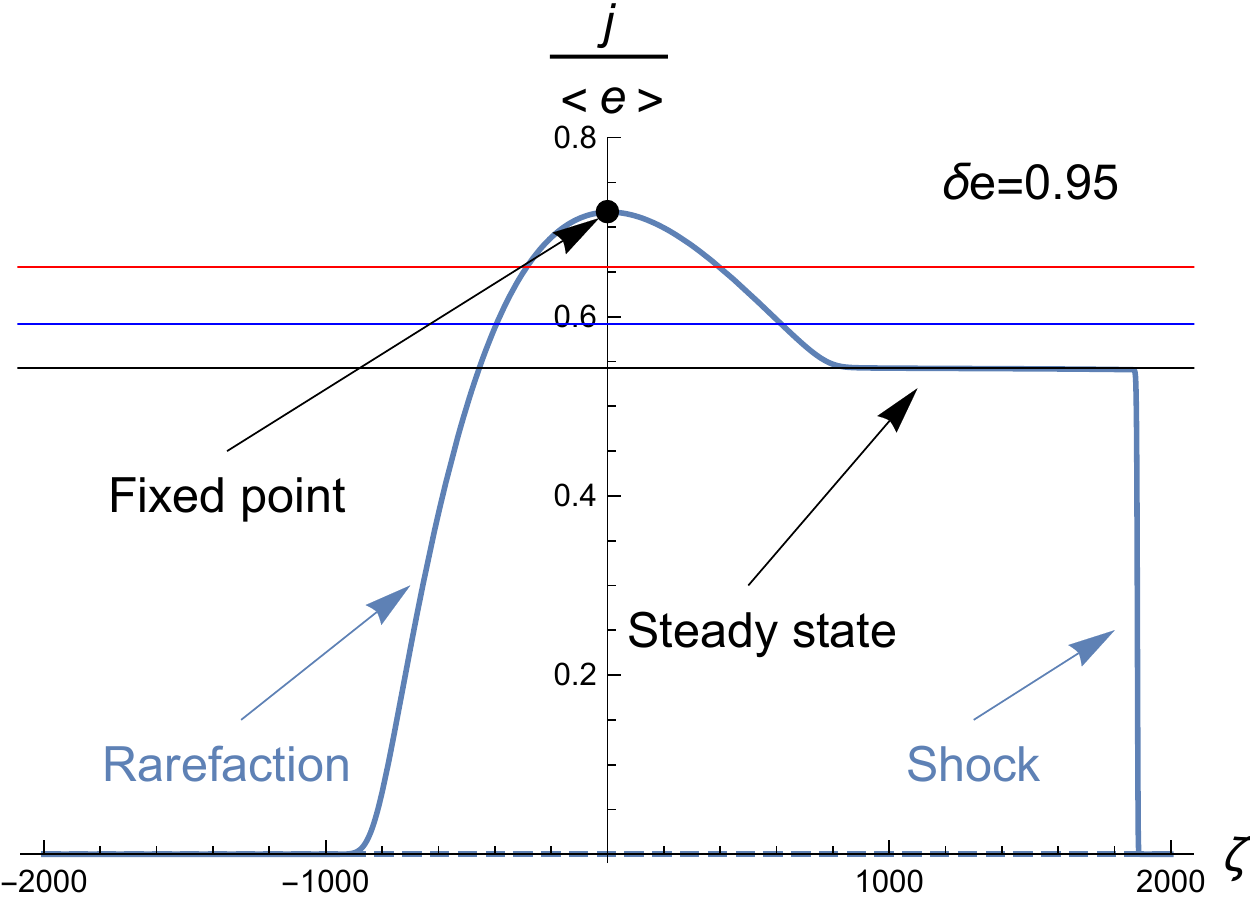}
\caption{A numerical solution to the Riemann problem. The plots were obtained starting with an initial condition \eqref{E:iniRS} with $L = 8000$, $c=300$ and $\langle j \rangle=0$. Only one half of the box, centered around the origin, is depicted. The dashed curve corresponds to values of $e$ and $j$ at $t=0$ while the solid curve corresponds to values of $e$ and $j$ at $t=800$. The black, red and blue horizontal lines correspond to the predicted near equilibrium steady state associated with a rarefaction wave and shock pair (c.f., equation \eqref{E:steadystate}), a bad shock and good shock pair (c.f., references \cite{Chang:2013gba,Bhaseen:2013ypa}), and a non thermodynamic shock pair (c.f., reference \cite{Chang:2013gba}) respectively. The fixed point associated with a rarefaction solution which exists for $\delta e \geq 0.7536\ldots$ is represented by a black dot.}
\label{fig:RSnumerics}
\end{figure}

\begin{figure}
%\centering
\includegraphics[width=3in]{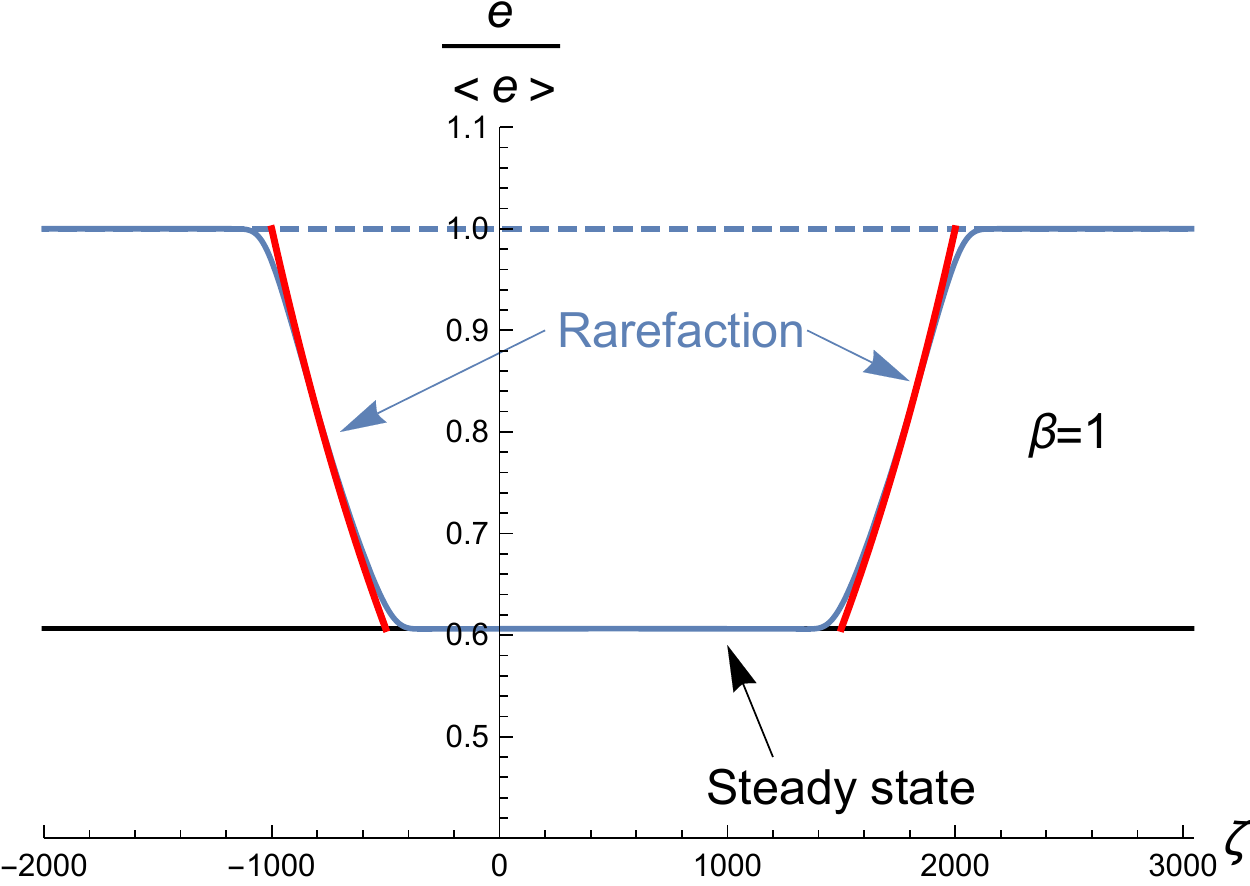}
\hfill
\includegraphics[width=3in]{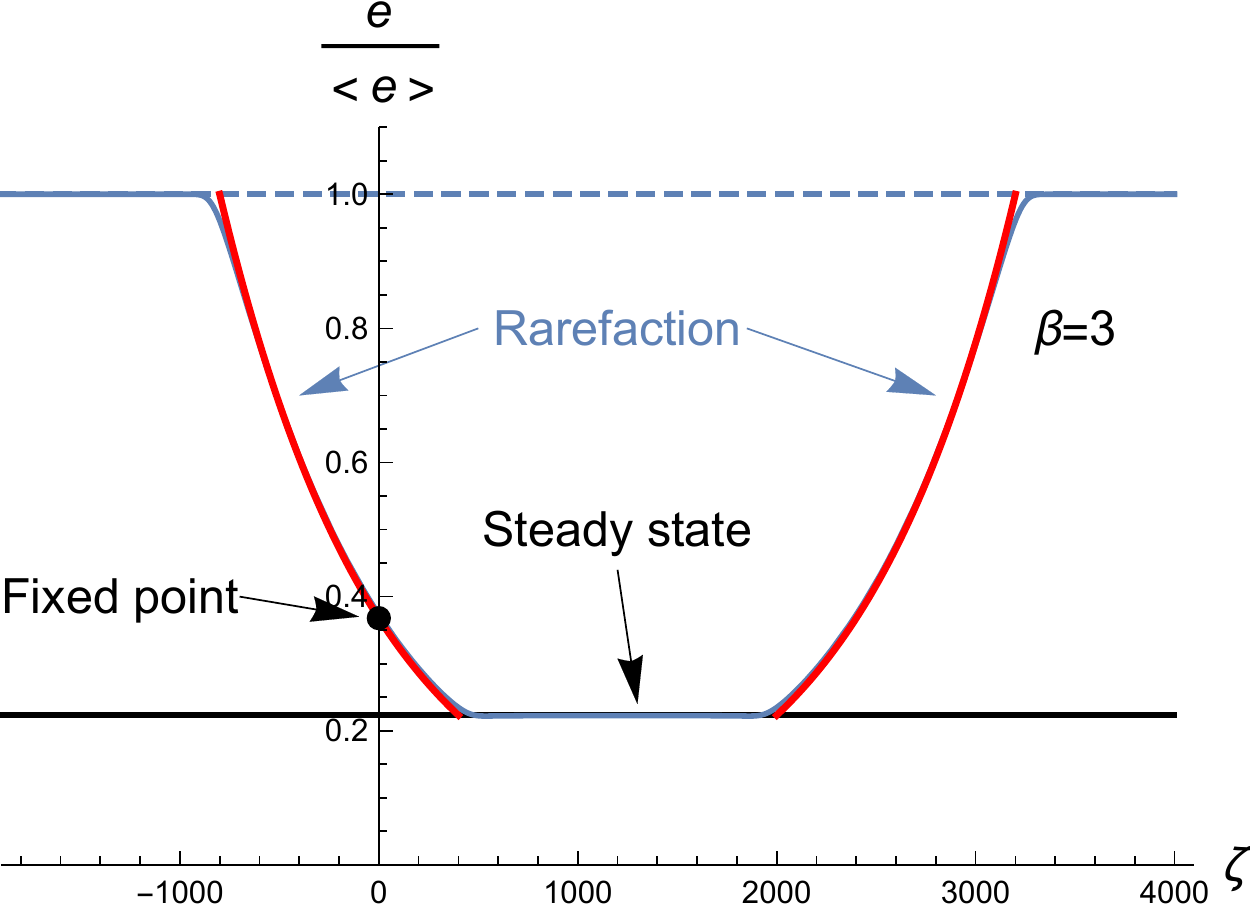}
\\
\includegraphics[width=3in]{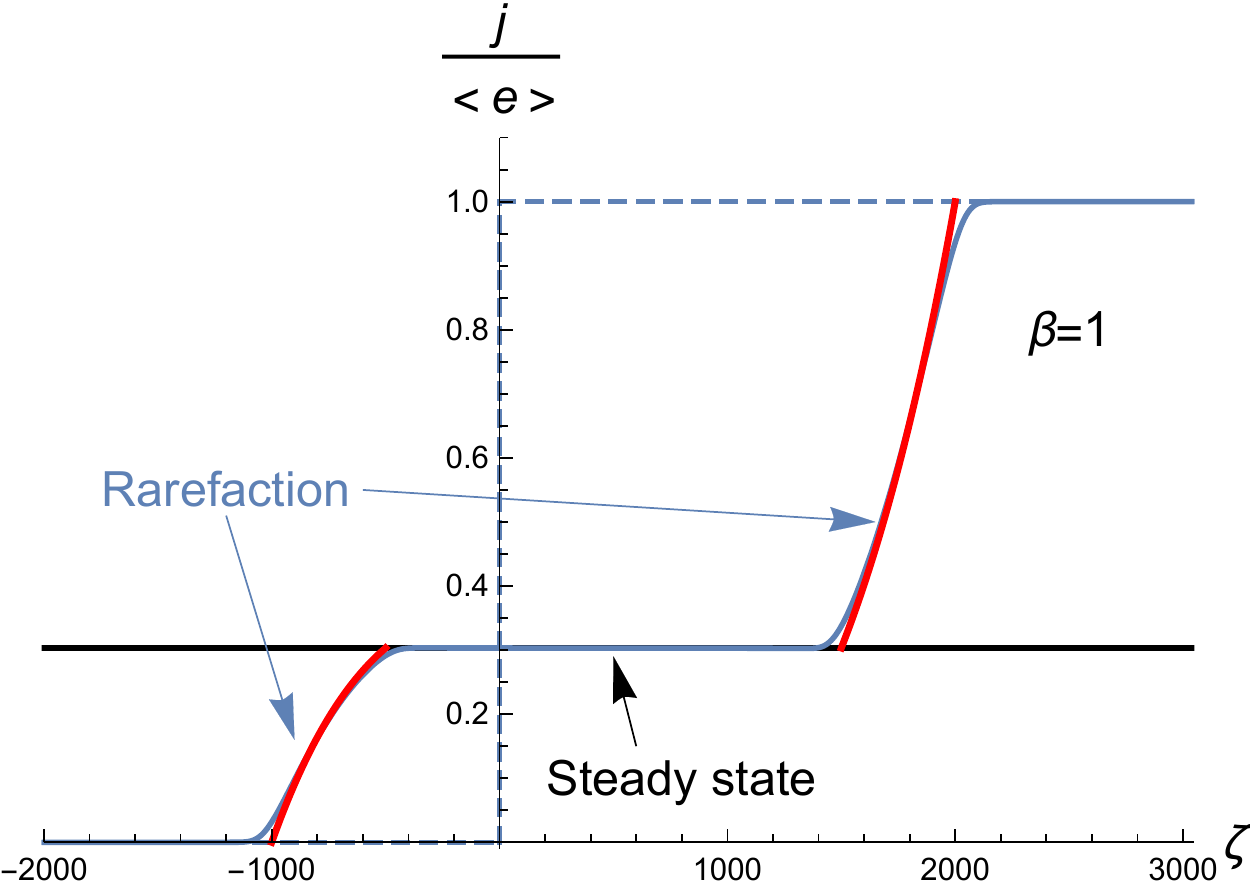}
\hfill
\includegraphics[width=3in]{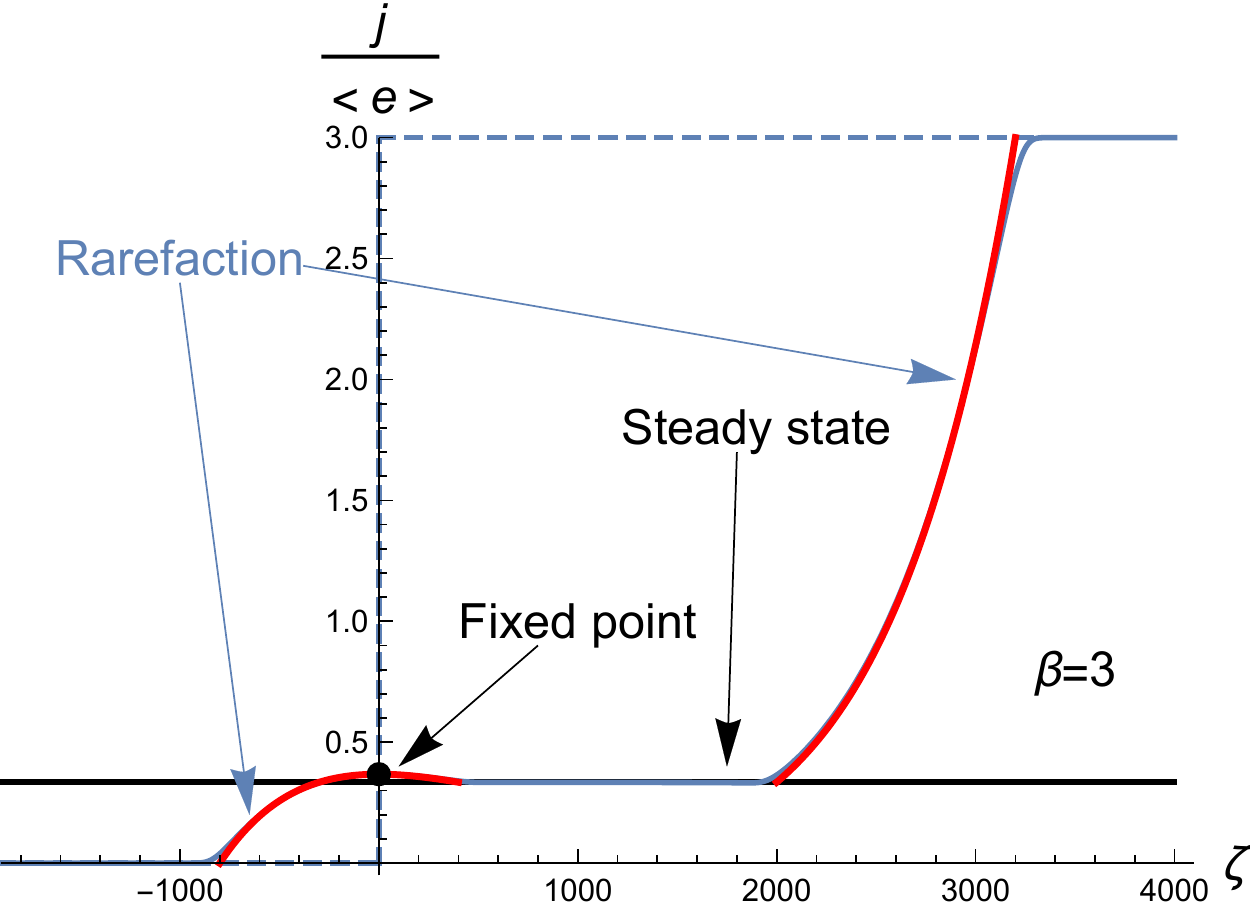}
\caption{Two numerical solutions to the Riemann problem in the RR case. The plots were obtained starting with a constant $e$ initial condition, $j_L = 0$, and fixed $\beta = j_R/e_L$, with $L = 8000$ and $c=200$. The dashed line corresponds to the solution at $t=0$ and the solid blue line at $t=1000$.  The solid red curves are the rarefaction waves in the ideal limit, without gradient corrections. The horizontal black line is the predicted steady state value.
}
\label{fig:RR}
\end{figure}

\begin{figure}
%\centering
\includegraphics[width=3in]{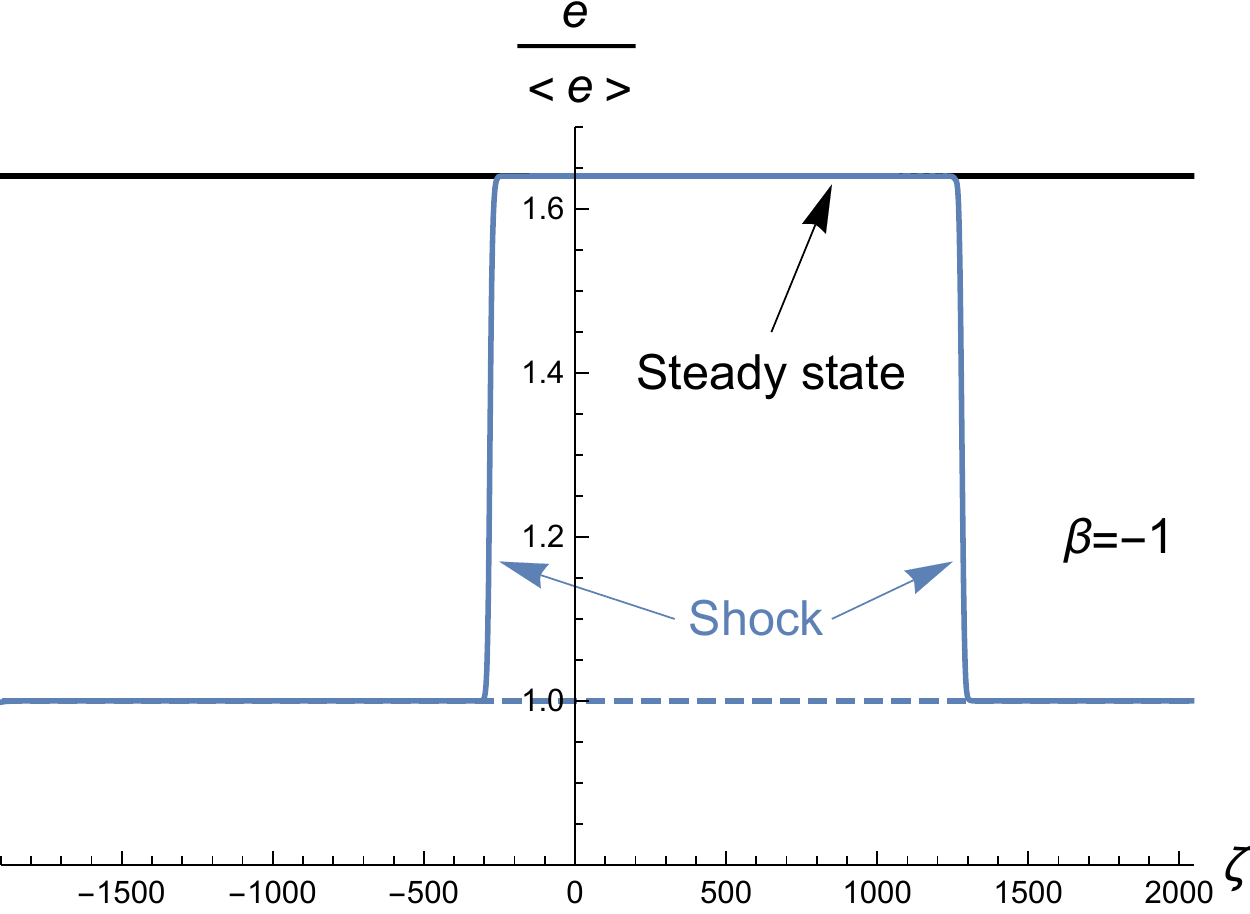}
\hfill
\includegraphics[width=3in]{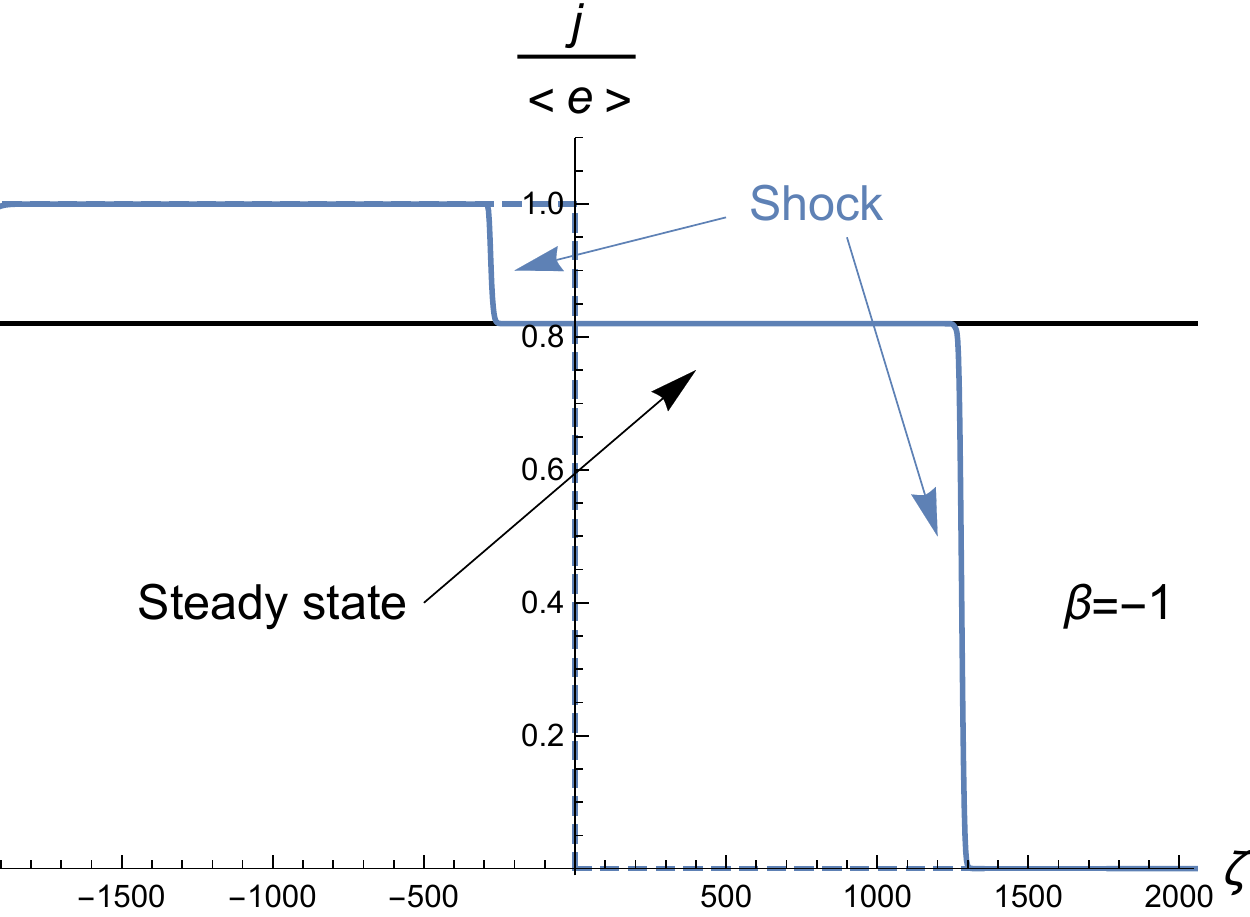}
\caption{A numerical solution to the Riemann problem in the SS case. The plots were obtained starting with a constant $e$ initial condition, $j_R = 0$, and $\beta = -j_L/e_L = -1$, with $L = 8000$ and $c=200$. The dashed line corresponds to the solution at $t=0$ and the solid blue line at $t=1000$.  The horizontal black line is the predicted steady state value.
}
\label{fig:SS}
\end{figure}

In figures \ref{fig:RSnumerics}, \ref{fig:RR}, and \ref{fig:SS}, we have plotted typical results for numerical solutions to \eqref{E:Tmn}, corresponding to RS, SS, and RR configurations. 
The resulting values of $e$ and $j$ seem to approach the predicted values of $e_0$ and $j_0$ at long times---at least as far as our numerical precision can be trusted (see appendix \ref{A:bestiary}).
In particular, in the RS case, we approach the steady state value \eqref{E:steadystate}; in the SS case, we approach \eqref{E:steadystateSS}; and in the RR case, we approach \eqref{E:steadystateRR}.
As we discuss in greater detail in the next section, one place where gradient effects show up and do not disappear as a function of time is in the shock width.  

One may speculate that the agreement between the predicted steady state in the absence of gradient corrections and the numerical results is associated to the fact that the gradient corrections, even though order one in our system of units, come with dimensionful  coefficients.  In the language of the renormalization group, they conform to irrelevant couplings.  Perhaps it is for this reason that at long enough time and in a large enough box, we may be able to ignore these corrections for the most part.

\subsection{Restoring gradient corrections}
\label{SS:restoring}

In this section, we try to gain a better handle over the gradient corrections and their affect on the predicted steady state values. 
The analysis here is incomplete and approximate.  To overcome the deficiencies of paper and pencil estimates, we include some numerical solutions to the conservation equations (\ref{E:Tmn}) that provide support for the estimates.
We will consider separately corrections to each of the features we found in the idealized limit: the steady state and asymptotic regions with constant $e$ and $j$, a shock wave, a rarefaction wave, and the discontinuity at the edge of the rarefaction.  

\subsubsection*{Corrections to constant regions}

Corrections to a constant $e$ and $j$ region are easiest to analyze.  Assuming the fluctuations are small, we look for linearized solutions of the form $e = e_0 + \delta e \, \exp(-i \omega t + i k \zeta)$ and $j= j_0 + \delta j \, \exp(-i \omega t + i k \zeta)$.  
We find two propagating modes
\begin{equation}
\omega = \left( \pm 1 + \frac{j_0}{e_0} \right) k - i k^2  \ .
\end{equation}
These two modes are damped sound modes whose speed is shifted by the fluid velocity $\beta = j/ e$.
The gradient corrections appear here in the form of the damping term $ik^2$ in the dispersion relation.  Given this result, we anticipate that we will be able to correct a constant $e$ and $j$ region by taking an appropriate linear superposition of sound waves.  The damping suggests that at long times the solution can only involve constant $e$ and constant $j$.

As a side comment, an odd thing about these mode relations is that they are exact.  Recall that in first order viscous hydro, we would typically solve an equation of the form $\omega^2 + i \Gamma k^2 \omega - k^2  = 0$ for $\omega$, in the case of vanishing background fluid velocity.  If this equation were treated as exact, the solutions for $\omega$ would be non linear in $k$ and therefore have higher order contributions, i.e.\ $O(k^3)$, $O(k^4)$, etc.,  when expanded around small $k$.

\subsubsection*{Corrections to shocks}

\begin{figure}
\centering
\includegraphics{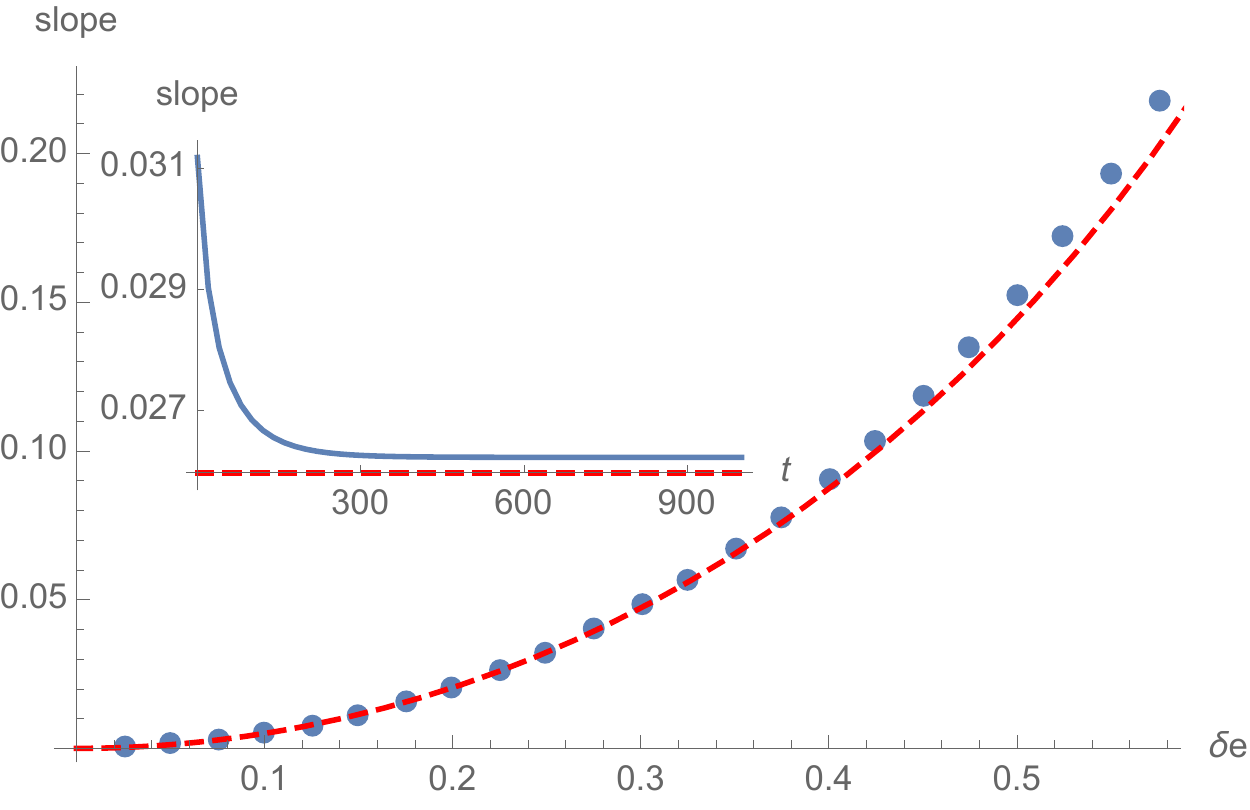}
\caption{ A numerical simulation of stationary shocks.  We start from an initial condition $e = \langle e\rangle(1 + \delta e \tanh (c \sin (2 \pi x / L) ))$, $j=1$ with parameters $L=8000$ and $c=1.2(L \delta e/4\pi)$.  We chose $e_r$ and $e_l$ to produce a stationary shock ($e_l =\frac{\sqrt{1-\text{$\delta $e}}}{\sqrt{\text{1+$\delta $e}}}$, $e_r =\frac{\sqrt{1+\delta e}}{\sqrt{1-\delta e}}$) using the RH relations. We then plot the value of the slope of the shock after the system has settled into a steady state.  This is compared with the weak shock solution (\ref{E:weakshock}), given by the dashed red line. The inset plot shows the relaxation from the initial conditions to the steady state for $\delta e = 0.23$.}
\label{fig:weakshockslope}
\end{figure}

The gradient corrections should act to smooth a shock and give it some characteristic width.
We estimate this width in a frame in which the shock is not moving, i.e.\ $s=0$. 
In this frame, $j_r = j_l$ and $e_r e_l = j_l^2$.  
We can find a solution for the shock profile in the case where the shock is weak $e_r  \sim e_l$:
\begin{align}
\label{E:weakshock}
	e&=\langle e \rangle \left[ 1 +  \delta e \tanh \left( \frac{\zeta \delta e}{2} \right)
- \frac{ \delta e^2}{2} \sech\left( \frac{ \zeta \delta e}{2} \right)^2 \log \cosh \left( \frac{\zeta \delta e}{2} \right) + O(\delta e^3)   \right] \,, \\
	j&=\langle j \rangle \left[1 \hspace{1.15in}+ \frac{\delta e^2}{2} \sech\left( \frac{ \zeta \delta e}{2} \right)^2 + O(\delta e^3)   \right] \ ,
\end{align}
where we have defined 
\[
\langle e \rangle \equiv \frac{e_r + e_l}{2}\,,
\quad
\delta e \equiv \frac{e_r -  e_l}{ e_r +  e_l}\,,
\quad
\hbox{and}
\quad
\langle j \rangle \equiv \frac{j_r + j_l}{2} \ .
\]
We can see in figure \ref{fig:weakshockslope} that even for values of $\delta e \sim 1/2$, that $\langle e \rangle \delta e^2 / 2$ appears to be a good estimate for the slope of the shock.\footnote{%
 We found that when $\delta e = 0.8$ the relative error between \eqref{E:weakshock} 
 and the numerical solution grew to $\sim 13\%$. As $\delta e$ gets closer to one numerical error is more difficult to control.
}
In appendix \ref{A:shockentropy}, we show that this shock profile produces, at the correct subleading order in a large $d$ expansion, the correct (positive) amount of entropy predicted by the RH relations.

\subsubsection*{Corrections to a rarefaction}

\begin{figure}
\centering
\includegraphics{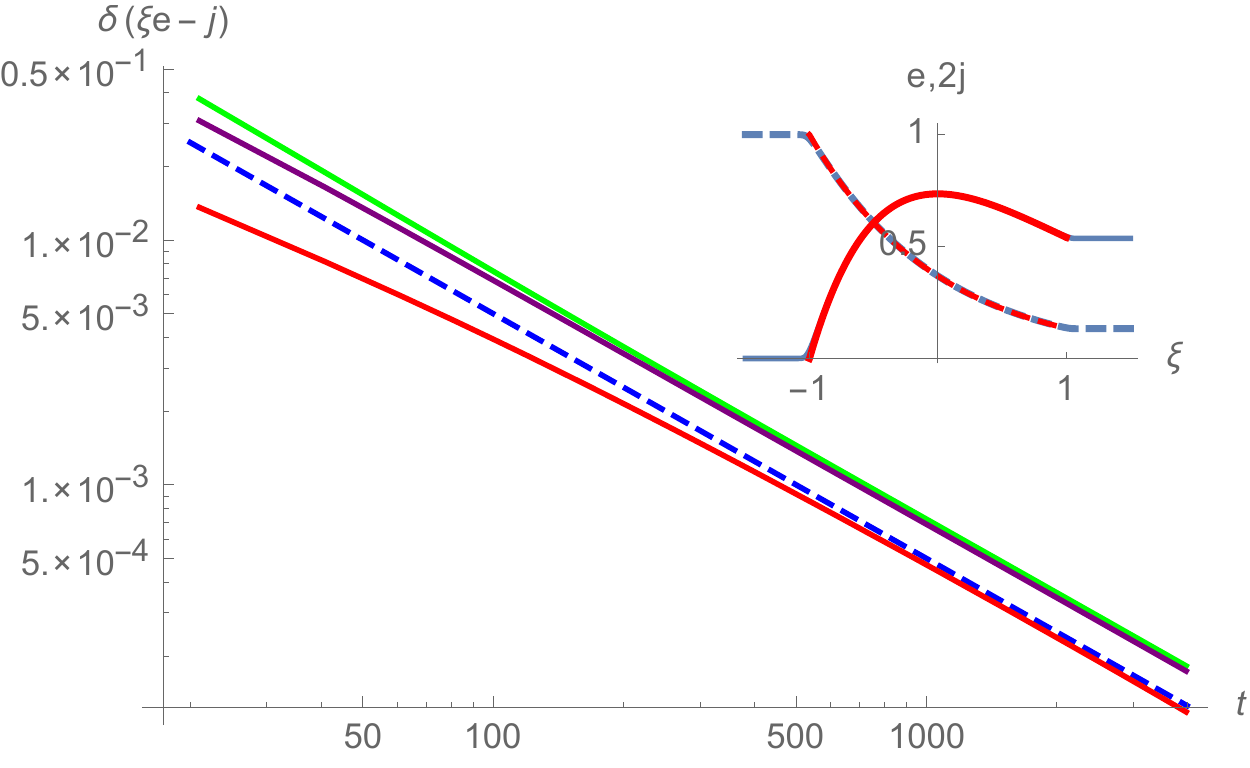}
\caption{
A plot of $\delta (\xi e - j)$ vs.\ time at three different points in a single rarefaction wave.  The quantity $\delta (\xi e - j)$ is the difference between the zeroth order prediction (\ref{E:rarefaction}) and numerics.  The rarefaction wave spreads from $\xi_l = -1$ to $\xi_r = 1$.  The three points correspond to $\xi = -1/2$ (red), $\xi= 0$ (purple) and $\xi = 1/2$ (green).   The dashed line $1/(2t)$ is a guide to the eye.  Inset: the rarefaction profile at $t = 3000$.  Dashed lines correspond to $e$ while the solid lines correspond to $j$.  The blue curve is numeric, while the red curve is the ideal result (\ref{E:rarefaction}).
}
\label{fig:xiej}
\end{figure}

We will perform two estimates of gradient corrections to the rarefaction wave.  The first estimate is a correction to the interior of the wave far from the edges where it joins onto constant $e$ and $j$ regions.  The second estimate is a correction to the discontinuity where the rarefaction joins a constant region.
For the first estimate, we assume an ansatz for the long time behavior of the rarefaction wave:
\begin{eqnarray*}
e &=& e_0(\xi) + \frac{\log t}{t} e_l(\xi) + \frac{1}{t} e_1(\xi) + O((\log t)^2/ t^2)\ , \\
j &=& j_0(\xi) + \frac{\log t}{t} j_l (\xi)+ \frac{1}{t} j_1(\xi) + O((\log t)^2/ t^2)\ ,
\end{eqnarray*}
where
\begin{eqnarray}
e_0 &=& c_1 \exp(\mp \xi) \ , \; \; \;
j_0 = (\pm 1+\xi) c_1 \exp(\mp \xi) \ , \\
e_l &=& 2 c_1 \exp(\mp \xi) - \frac{1}{2} c_2 \exp(\mp \xi/2) \ , \; \; \;
j_l = \xi e_l \ , \\
j_1 &=& \pm \exp(\mp \xi) (c_1 - c_2 \exp(\pm \xi/2)) +\xi e_1 \ .
\end{eqnarray}
With an appropriate choice for the integration constant $c_1$, the expressions for $e_0$ and $j_0$ become the same as we had before (\ref{E:rarefaction}).  There are subleading corrections that scale as $1/t$ and $\log(t) / t$ that depend on a second integration constant $c_2$ and an arbitrary function $e_1(\xi)$, both presumably set by the initial conditions.  Note that the combination $\xi e - j$ is independent of the arbitrary function $e_1(\xi)$ at order $1/t$. In figure \ref{fig:xiej}, the numerics confirm that the corrections to $\xi e - j$ do indeed scale as $1/t$.

\begin{figure}
\centering
\includegraphics{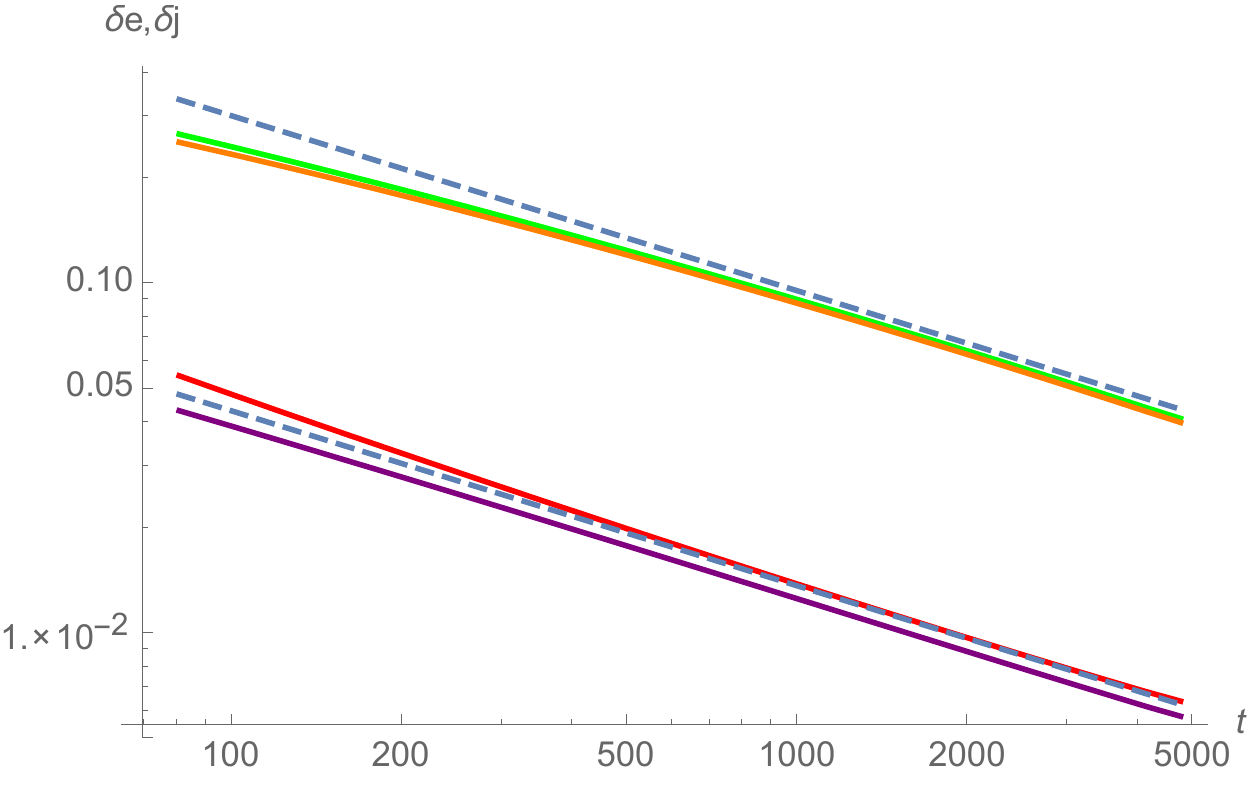}
\caption{
A log log plot of $\delta e$,$\delta j$ vs.\ time at the endpoints of a rarefaction wave, where $\delta e = e-e_0$ and $\delta j = j-j_0$ and $e_0$ and $j_0$ are from the zeroth order prediction (\ref{E:rarefaction}).  As in figure \ref{fig:xiej} the rarefaction wave spreads from $\xi_l = -1$ to $\xi_r = 1$.  The four curves correspond to $e(1)$ (red), $e(-1)$ (purple) and $j(1)$ (green) and $j(-1)$ (orange).   The dashed lines $0.43 t^{-1/2}$ and $3 t^{-1/2}$ are a guide to the eye.
  }
\label{fig:endpoints}
\end{figure}

Last, we would like to heal the discontinuity at the edge of a rarefaction wave. The $\tanh$ function we found above heals the discontinuity in the shock case, making the question of what happens at the edge of a shock less pressing.  Consider a case where the rarefaction wave meets a steady state at $\zeta=0$, with the rarefaction region to the right and the steady state to the left.  (We can always move the meeting point away from $\zeta=0$ by boosting the solution $\zeta \to \zeta + vt$.)  
With the intuition that the second order gradients in the conservation equations are dominant and render the behavior similar to that of a heat equation with $1/\sqrt{t}$ broadening, we look for an approximate late time solution of the form
\begin{eqnarray}
e &=& e_0 + \frac{1}{\sqrt{t}} e_1(\chi) + O(t^{-1})\ , \\
j &=& j_0 + \frac{j_1}{\sqrt{t}}+  \frac{1}{t} j_2(\chi) + O(t^{-3/2}) \ ,
\label{E:endpoints}
\end{eqnarray}
defining $\chi \equiv \zeta / \sqrt{t}$.
We find that $j_0 = \pm e_0$, that $j_1$ is constant, and that  
\[
j_2'(\chi) = \mp \frac{e_1(\chi) e_1'(\chi) }{ e_0} + \left( 4 \frac{e_1'}{e_0}\pm 1 \right) j_1 \ .
\]
Note that the relation $j_0 = \pm e_0$ is consistent with a rarefaction meeting a steady state region at $\zeta=0$.  
These relations for the $j_i$ lead
to a second order, nonlinear differential equation for $e_1$:
\begin{equation}
e_1'' +\left( \frac{\chi}{2}+ \frac{\pm e_1 - j_1}{e_0} \right) e_1'  + \frac{e_1}{2}  \mp \frac{j_1}{4} = 0 \ .
\end{equation}
Remarkably, this equation can be written as a total derivative and integrated to yield
\begin{equation}
\label{e1first}
\pm \frac{e_1^2}{2 e_0} + e^{-\chi^2/4} \partial_\chi ( e^{\chi^2/4} e_1) - \frac{j_1 e_1}{e_0} \mp \frac{j_1}{4} \chi = c_1 \ ,
\end{equation}
where $c_1$ is another integration constant.  
The integration constants reflect a translation symmetry of both $e_1$ and $\chi$.  We can shift $\chi \to \chi + j_1/e_0$ and
$e_1(\chi) \to e_1( \chi - j_1/e_0) \pm j_1/2$.  The shifts send $j_1 \to 0$ and $c_1 \to c_1 \mp 3 j_1^2 / 8 e_0$ in the equation (\ref{e1first}).  
If we apply the boundary condition that both $e_1(\chi)$ and $e_1'(\chi)$ vanish in the steady state region $\chi \to -\infty$, then we must set $c_1=0$, and the resulting first order differential equation becomes separable.
To match onto the rarefaction region, we require that $e_1' \to \pm e_0$ as $\chi \to \infty$. 
This boundary condition fixes the remaining integration constant associated with the first order equation (\ref{e1first}), and the solution for $e_1$ is then
\begin{equation}
e_1 = \pm \frac{2 e_0 e^{-\chi^2/4}}{\sqrt{\pi} \erfc(\chi/2)} \ .
\end{equation}
As we choose the rarefaction region to match onto the steady state at $\chi=0$, we conclude that the integration constant $j_1$ in the original differential equation must be zero as well. We can check numerically that a $1/\sqrt{t}$ scaling is consistent with the behavior at the endpoints of a rarefaction solution. See figure \ref{fig:endpoints}.

\section{Discussion}
\label{sec:discussion}

We presented a solution to the Riemann problem for the conservation equations (\ref{E:Tmn}).  Through fluid-gravity and the AdS/CFT correspondence, these equations describe, in a large $d$ limit, both the  dynamics of a black hole horizon and also the dynamics of a strongly interacting conformal field theory.  

There are a number of possible future directions for research.  The simplest is perhaps to include a transverse velocity. With a transverse velocity, in addition to the shock and rarefaction waves, there will in general be a contact discontinuity \cite{Taub, Thompson, MachPietka,Spillane:2015daa}.  It is known (and perhaps intuitive given the similarity to a counter flow experiment), that the contact discontinuity is in general unstable to the development of turbulence \cite{Mach}.  It would be interesting to see what precisely happens in our large $d$ limit.  Another more complicated extension is the inclusion of a conserved charge.  The large $d$ equations of motion in the presence of a conserved charge are available from ref.\ \cite{Emparan:2016sjk}.  Once again, a contact discontinuity is expected (see for example \cite{Spillane:2015daa}) although whether such a discontinuity is stable or unstable to turbulence is unclear.  More ambitiously, one could consider what happens for the holographic dual of a superfluid or superconductor \cite{Romero-Bermudez:2015bma,Emparan:2013oza,Gubser:2008px,Hartnoll:2008vx,Hartnoll:2008kx,Herzog:2008he,Herzog:2011ec}.

Another possible direction is the addition of higher curvature terms to the dual gravitational description.  One could presumably  tune the $d$ dependence of these terms such that higher order gradient corrections appear in the conservation equations (\ref{E:Tmn}) and also such that the first and second order transport coefficients are tuned away from the values examined in this paper.

Perhaps the most interesting direction for future study is the connection to black hole dynamics.  What can we learn about black holes through the connection to hydrodynamics in a large $d$ limit?

\section*{Acknowledgments}

We would like to thank S.~Bhattacharyya, S.~Cremonini, J.~Glimm, V.~Hubeny, D.~Huse, A.~Lucas, A.~Ori, M.~Rangamani, and K.~Schalm for discussion.
M.~S. and C.~P.~H.\ were supported in part by NSF Grant No.~PHY13-16617.  A.~Y.\ was supported by the ISF under grant numbers 495/11, 630/14 and 1981/14, by the BSF under grant number 2014350, by the European commission FP7, under IRG 908049 and by the GIF under grant number 1156/2011.

\begin{appendix}

\section{Comment About Entropy Production Across a Shock}
\label{A:shockentropy}

In the ideal limit, in addition to conservation of energy and momentum, we can write down a conservation condition for the entropy current, 
$\partial_\mu \tilde{J}_S^\mu = 0$ where 
\begin{equation}
\label{E:JSideal}
	\tilde{J}_S^\mu = (\epsilon + p) u^\mu /T.
\end{equation}
This conservation condition would naively seem to lead to an additional Rankine-Hugoniot relation across a single shock. 
As is well known in the hydrodynamics community (see for example \cite{Lucas:2015hnv}), since shocks create entropy this third Rankine-Hugoniot relation is violated. 
Let us parameterize a possible violation of the additional Rankine-Hugoniot relation by $\Delta$.
\begin{equation}
\Delta = s [ \tilde{J}_S^t ] - [ \tilde{J}_S^\zeta ]  
\end{equation}
where the square brackets are the same as those in \eqref{E:RHeqns}.
One finds
\begin{equation}
\label{sprodstationary}
\Delta = \frac{2\pi }{\sqrt{e_r e_l} d^2} \left( e_r^2 - e_l^2 - 2 e_r e_l \log \left( \frac{e_r}{e_l} \right) \right) + O(d^{-3}) \ .
\end{equation}
Equation \eqref{sprodstationary} can be obtained by using a large $d$ expression for the entropy current \eqref{E:JSgravity} along with the Rankine-Hugoniot relations for energy and momentum, \eqref{E:RHeqns} supplemented by \eqref{E:fullTmn} and \eqref{E:fullT2mn}.  Note that in the asymptotic regions, the gradient terms will all vanish.
(It is also possible to start with a finite $d$ result, using for example refs.\  \cite{Lucas:2015hnv}  or \cite{Spillane:2015daa}, and then take a large $d$ limit directly.)

The non-conservation of entropy \eqref{sprodstationary} can be captured by the leading viscous corrections to the shock width \eqref{E:weakshock} when the energy difference is small. Indeed, using \eqref{E:divJgravity}
\begin{equation}
\partial_\mu \tilde{J}_S^\mu = \frac{8 \pi }{d^2}  \frac{j_0^2 (e')^2}{e^3}+ O(d^{-3}) = \frac{2 \pi  j_0^2 \delta e^4}{ d^2 \langle e \rangle} \sech \left(\frac{\zeta \delta e}{2} \right)^4+ O(\delta e^5, d^{-3}) \ .
\end{equation}
Integrating this divergence over the $\zeta$ direction leads to
\begin{equation}
\label{Eproduction}
\int \partial_\mu \tilde{J}_S^\mu \d \zeta = \frac{16 \pi \langle e \rangle \delta e^3}{3 d^2 } + O(\delta e^4, d^{-3}) \ ,
\end{equation}
which agrees with a small $\delta e$ expansion of (\ref{sprodstationary}).

\section{A bestiary of plots}
\label{A:bestiary}
In section \ref{SS:numerical} we studied the numerical solutions to the Riemann problem for various initial energy and velocity profiles associated with $RR$, $RS$ and $SS$ type solutions. In what follows we provide additional evidence that at late times the full numerical solution to the Riemann problem approaches the appropriate predicted steady state values $e_0$ and $j_0$ and fixed point values $e_s$ and $j_s$.

\subsection{RR configurations}
\label{SS:RRc}
\begin{figure}
\centering
\includegraphics[width=3 in]{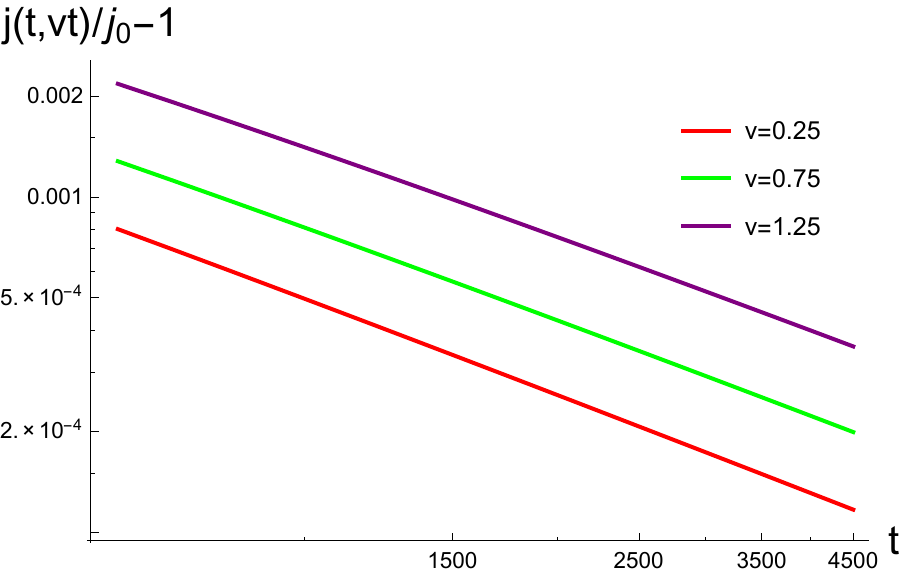}
\hfill
\includegraphics[width=3 in]{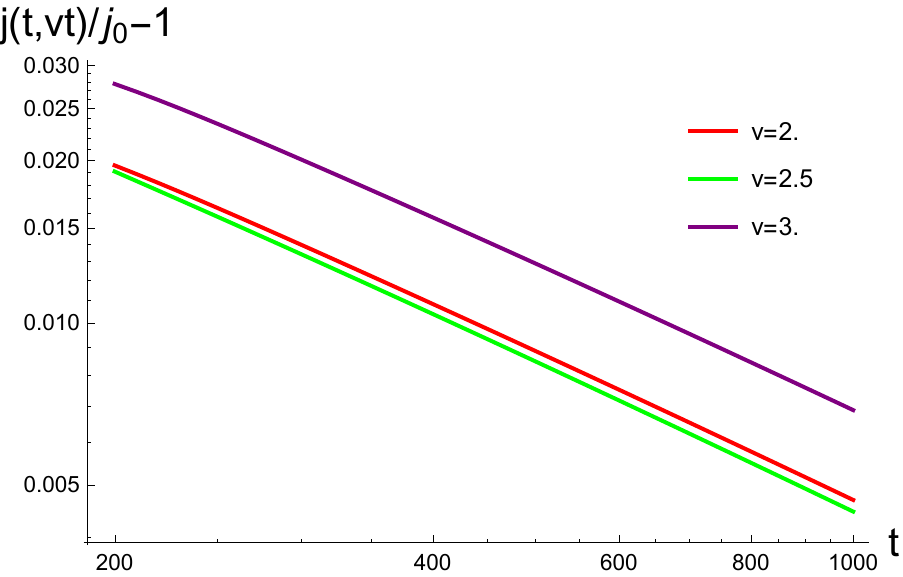} \\
\includegraphics[width=3 in]{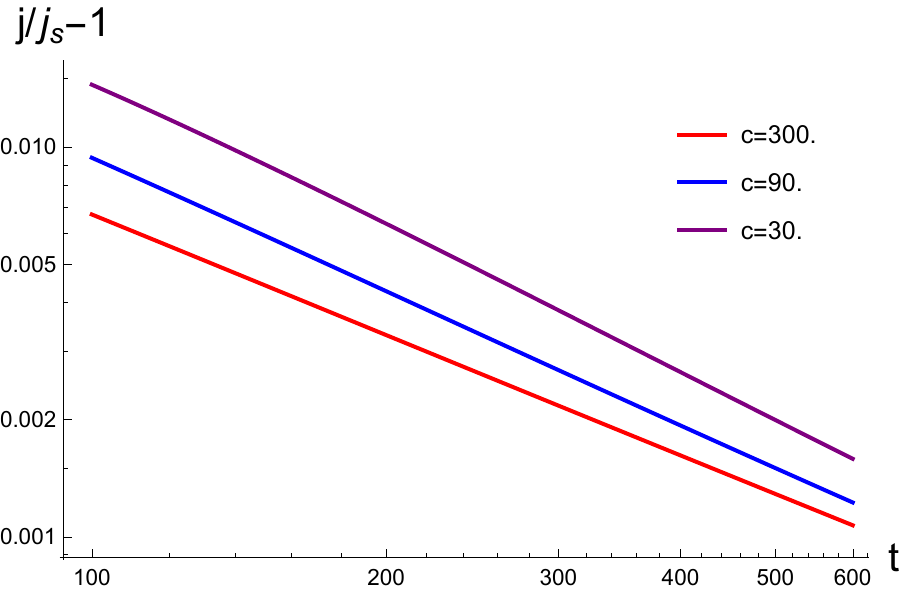}
\hfill
\includegraphics[width=3 in]{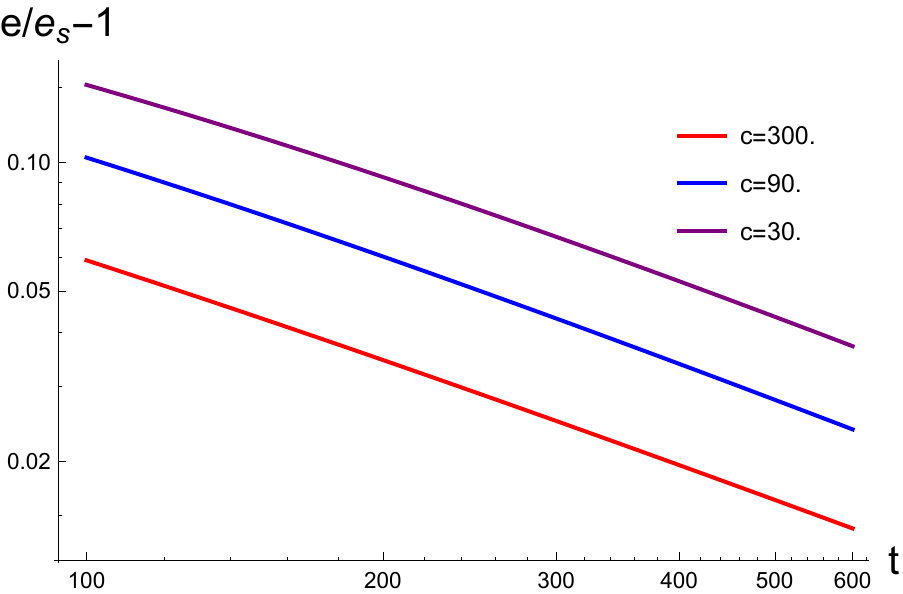}
\caption{
Late time behavior of the steady state and fixed point for RR type configurations. Top plots: The deviation of $j(t,\zeta=vt)$ from the predicted steady state value $j_0$ for various values of $v$.  The initial conditions are given by \eqref{E:iniRR} with $L=20000$, $\ell=8000$, and $c=300$ and $j_*=1.8$ for the top left plot and $L=8000$, $\ell=2000$, $c=100$ and $j_*=5$ for the top right plot. Both the results roughly fit a $\sim t^{\alpha}$ asymptotic behavior with $\alpha \sim 0.9$. Bottom plots: The deviation of $e$ and $j$ from the predicted fixed point value at $\zeta=0$ for various values of $c$. The initial conditions are given by \eqref{E:iniRR} with $L=16000$, $\ell=4000$ and $j_*=3$. Both the time dependence of $e/e_s-1$ and $j/j_s-1$ can be fit to a power law, $\sim t^{\alpha}$. For the energy density one finds that $\alpha$ gradually increases to $\alpha \sim 0.8$ as one approaches $c=300$. For the energy current $\alpha$ decreases to $\alpha \sim 1.1$ at $c=300$.
}
\label{fig:RRasymptotics}
\end{figure}

To generate an RR configuration we used the initial data
\begin{equation}
\label{E:iniRR}
	e = 1\,, \qquad
	j = \begin{cases} 
		f(\zeta) & 0 \leq \zeta < \ell/4 \\
		0 & \ell/4 \leq \zeta < L/2 - \ell/4 \\
		f(\zeta - L/2 - \ell/2) & L/2 - \ell/4 \leq \zeta < L/2+\ell/4 \\
		j_* & L/2 + \ell/4 \leq \zeta < L-\ell \\
		f(\zeta-L) & L-\ell \leq \zeta < L
	\end{cases}
\end{equation}
where
\begin{equation}
\label{E:fRR}
	f(\zeta) = \frac{1}{2} j_*\left(1-\tanh\left( c \sin \left(\frac{2\pi \zeta}{\ell} \right) \right) \right)\,.
\end{equation}
The analysis of section \ref{SS:implicit} predicts a steady state of the form
\begin{equation}
	e_0 = \exp\left(-j_*/2\right)
	\qquad
	j_0 = \frac{j_*}{2}\exp\left(-j_*/2\right)\,.
\end{equation}
Once $j_*\geq 2$ one should find a fixed point with $e_s = j_s = \exp(-1)$. We find that the numerical solution approaches the predicted states via power law behavior, see figure \ref{fig:RRasymptotics}.

\subsection{SS configurations}
\label{SS:SSc}
\begin{figure}
\centering
\includegraphics[width=3 in]{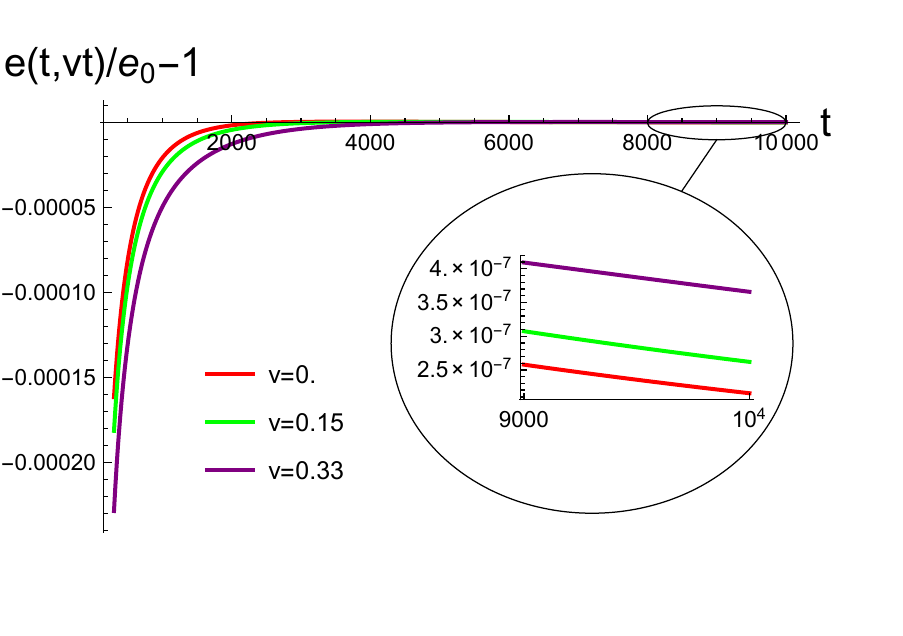}
\hfill
\includegraphics[width=3 in]{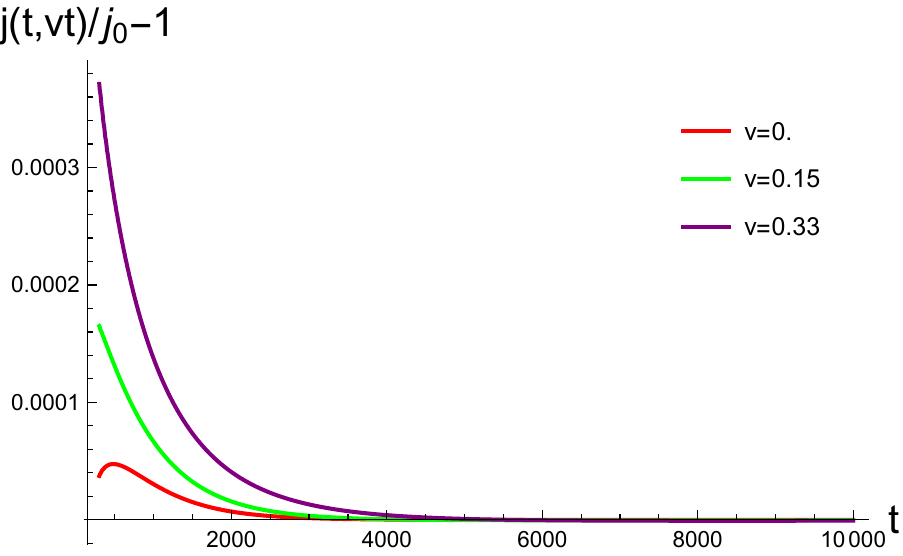} \\
\includegraphics[width=3 in]{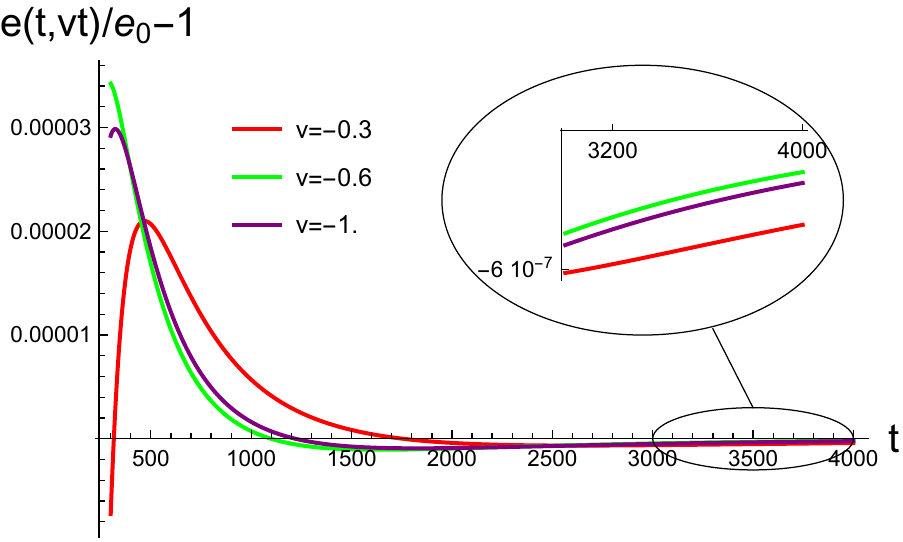}
\hfill
\includegraphics[width=3 in]{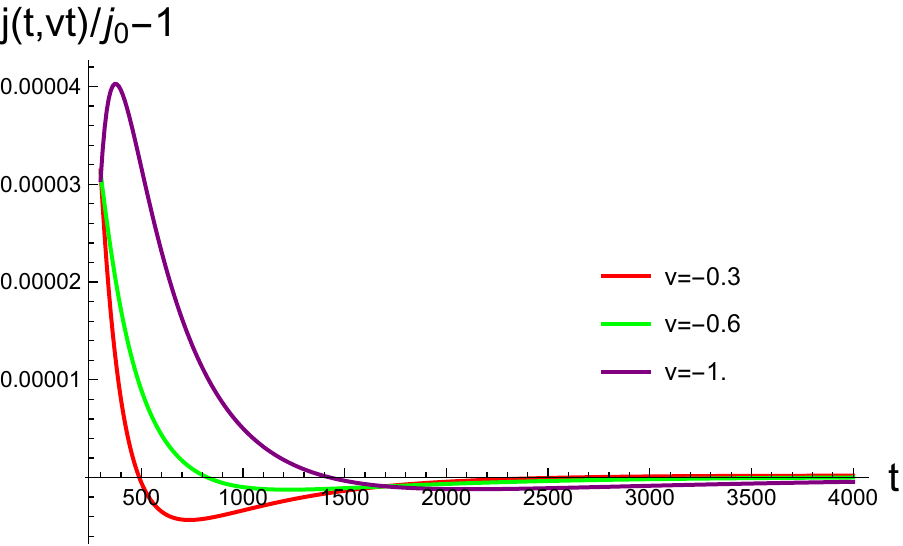}
\caption{
Late time behavior of the steady state and fixed point for SS type configurations. The plots show the deviation of $e(t,\zeta=vt)$ and $j(t,\zeta=vt)$ from the predicted steady state values $e_0$ and $j_0$ for various values of $v$.  The initial conditions are given by \eqref{E:iniRR} with $L=40000$ (top) or $L=20000$ (bottom), $\ell=2000$ and $c=100$. The top plots correspond to $j_*=-0.5$ and the bottom ones to $j_*=-2$. We expect that numerical error is of order $10^{-7}-10^{-8}$.
}
\label{fig:SSasymptotics}
\end{figure}

To generate an SS configuration we used the initial data \eqref{E:iniRR} with $j_*<0$.
The analysis of section \ref{SS:implicit} predicts a steady state of the form
\begin{equation}
	e_0 = \frac{1}{8} (8 + j_*^2 - j_* \sqrt{16+j_*^2}) \ , \quad 
	\frac{j_0}{e_0} = \frac{j_*}{2} \ .
\end{equation}
See figure \ref{fig:SSasymptotics} for a comparison with the numerical data.

\subsection{RS configurations}
\label{SS:RSc}
\begin{figure}
\centering
\includegraphics[width=3 in]{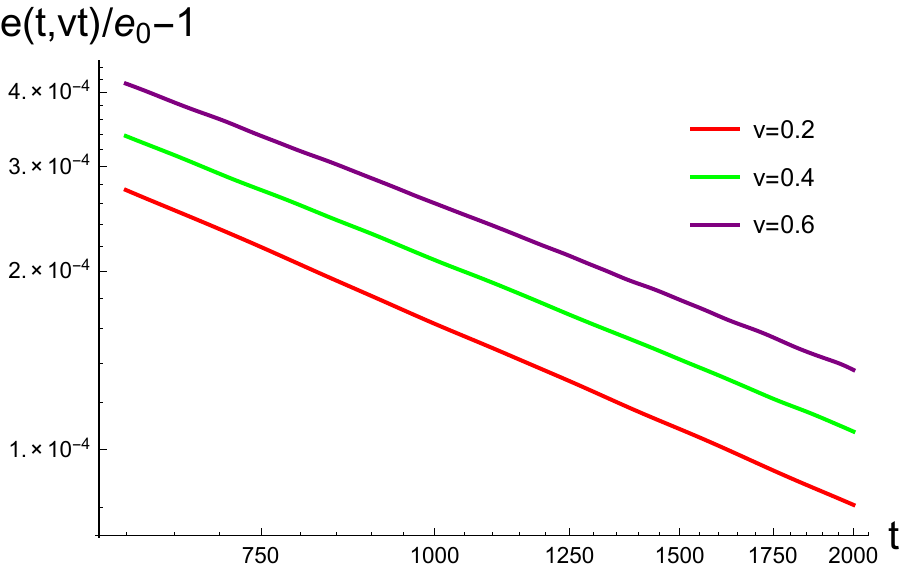}
\hfill
\includegraphics[width=3 in]{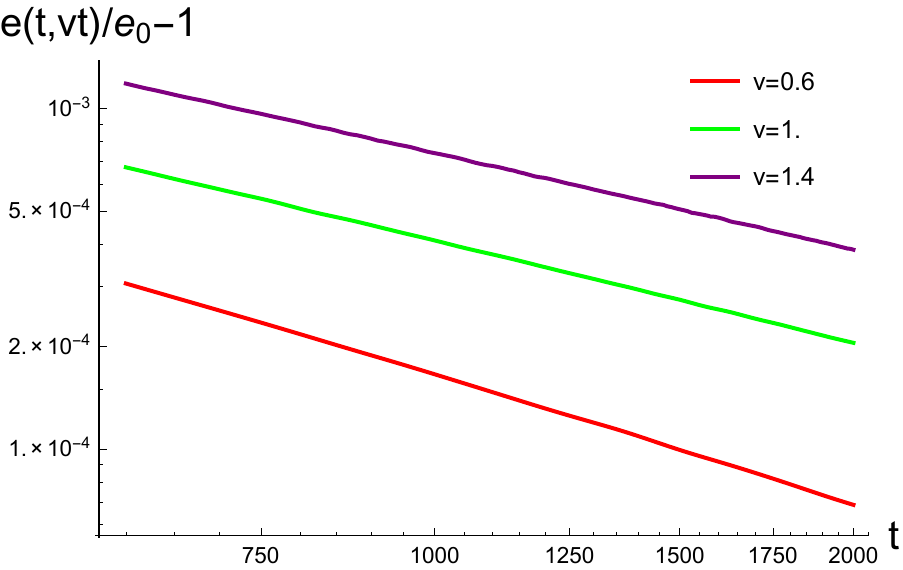} \\
\includegraphics[width=3 in]{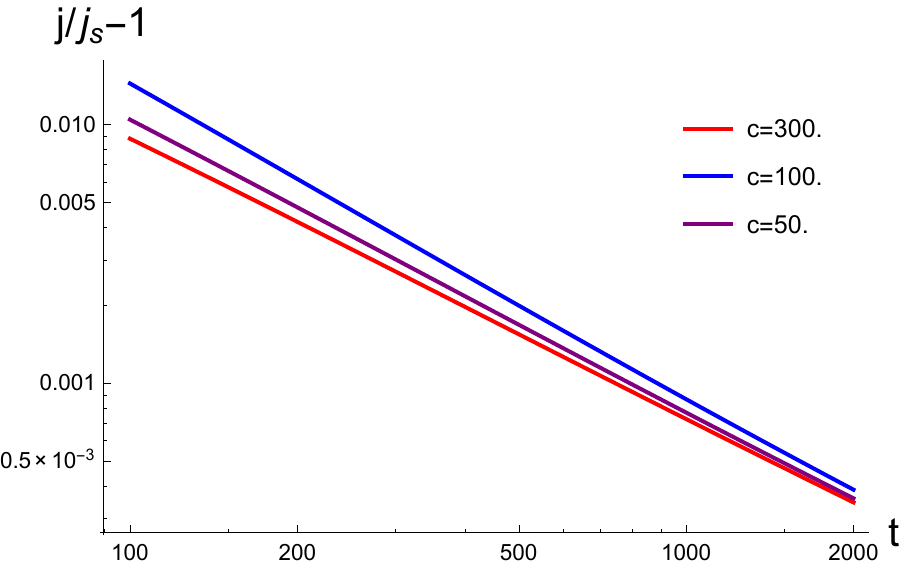}
\hfill
\includegraphics[width=3 in]{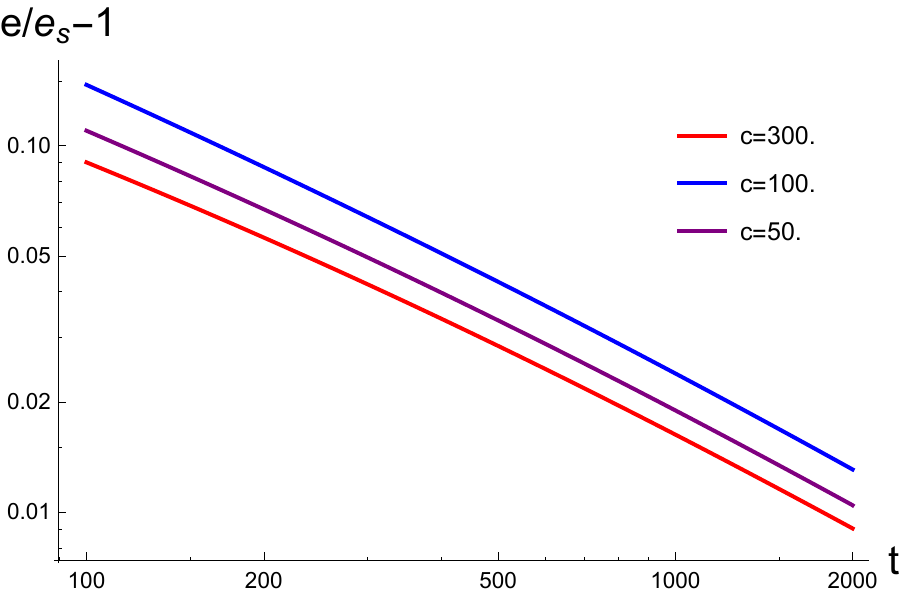}
\caption{
Late time behavior of the steady state and fixed point for RS type configurations. Top plots: The deviation of $e(t,\zeta=vt)$ from the predicted steady state value $e_0$ for various values of $v$.  The initial conditions are given by \eqref{E:iniRSv2} with $L=16000$, $\ell=2000$, and $c=100$ and $e_*=4$ for the top left plot and $e_*=9$ for the top right plot. Bottom plots: The deviation of $e$ and $j$ from the predicted fixed point value at $\zeta=0$ for various values of $c$. The initial conditions are given by \eqref{E:iniRR} with $L=16000$, $\ell=4000$ and $e_*=9$. Both the time dependence of $e/e_s-1$ and $j/j_s-1$ can be fit to a power law, $\sim t^{\alpha}$. For the energy density one finds $\alpha\sim 0.77$. For the energy current $\alpha \sim 1.1$.
}
\label{fig:RSasymptotics}
\end{figure}

To generate an RS configuration we used the initial data
\begin{equation}
\label{E:iniRSv2}
	j = 0\,, \qquad
	e = \begin{cases} 
		f(\zeta) & 0 \leq \zeta < \ell/4 \\
		e_* & \ell/4 \leq \zeta < L/2 - \ell/4 \\
		f(\zeta - L/2 - \ell/2) & L/2 - \ell/4 \leq \zeta < L/2+\ell/4 \\
		1 & L/2 + \ell/4 \leq \zeta < L-\ell \\
		f(\zeta-L) & L-\ell \leq \zeta < L
	\end{cases}
\end{equation}
where
\begin{equation}
\label{E:fRS}
	f(\zeta) = \frac{1}{2} \left(1+e_*\right) + \frac{1}{2}\left(e_*-1\right)  \tanh\left( c \sin \left(\frac{2\pi \zeta}{\ell} \right) \right) +e_*\,.
\end{equation}
The analysis of section \ref{SS:implicit} predicts a steady state of the form
\begin{equation}
	e_0 = s^2 \ , \quad 
	j_0 = s(s^2-1) \ .
\end{equation}
with 
\begin{equation}
	0 = \frac{1}{s} - s - \log\left(\frac{s^2}{e_*}\right) \ .
\end{equation}
According to the same analysis, once $e_* \geq \left(\frac{1+\sqrt{5}}{2}\right)^2 \exp(1)$ we will obtain a fixed point at the origin with $e_s = j_s = \exp(-1)$. An analysis of the late time behavior of the numerical solution can be found in figure \ref{fig:RSasymptotics}.

\subsection{Error analysis}
\label{SS:error}
In sections \ref{SS:RRc} and \ref{SS:RSc} we have fit the late time approach of the data to the predicted steady state and (or) fixed point values to a power law behavior. The fit was done using Mathematica's NonLinearModelFit routine \cite{Mathematica}. In detail, the late time data was discretized into order 1 time steps which were then fit to a $a/t^{\alpha}$ curve with $a$ and $\alpha$ as parameters. The standard errors for the fit were usually of order $10^{-3}$ to $10^{-4}$. Fits involving very small values of the slope parameter $c$ in \eqref{E:fRR} and \eqref{E:fRS} (c.f., the bottom plots of figures \ref{fig:RRasymptotics} and \ref{fig:RSasymptotics}) often had large standard errors.

\end{appendix}

\bibliographystyle{JHEP}
\bibliography{larged}

\end{document}